\documentclass[journal,nolists]{asmejour}
\usepackage[utf8]{inputenc}
\usepackage{accents}
\usepackage[mathscr]{euscript}
\usepackage{mathrsfs}
\usepackage{mathtools}
\usepackage{textcomp, gensymb}
\usepackage{booktabs}
\usepackage{multirow}
\usepackage{array}
\usepackage{algorithm, algpseudocode}
\usepackage{comment}

\usepackage{xcolor}

\usepackage[capitalize]{cleveref}

\hypersetup{%
	pdfauthor={Audrey Blizard},
	pdftitle={Journal Modeling for DHN},
}

\JourName{Dynamic Systems,\\ Measurement and Control}

\begin{document}

\SetAuthorBlock{Audrey Blizard\CorrespondingAuthor}{Department of Mechanical Engineering,\\
   The Ohio State University,\\
   Columbus, OH, 43210 \\
   email: blizard.1@osu.edu} 
\SetAuthorBlock{Stephanie Stockar}{Department of Mechanical Engineering,\\
The Ohio State University,\\
Columbus, OH, 43210 \\
email: stockar.1@osu.edu} 

\title{A Graph-Based Technique for the Automated Control-Oriented Modeling of District Heating Networks}

\begin{abstract}
Advanced control strategies for delivering heat to users in a district heating network have the potential to improve performance and reduce wasted energy. To enable the design of such controllers, this paper proposes an automated plant modeling framework that captures the relevant system dynamics, while being adaptable to any network configuration. Starting from the network topology and system parameters, the developed algorithm generates a state-space model of the system, relying on a graph-based technique to facilitate the combination of component models into a full network model. 
The accuracy of the approach is validated against experimental data collected from a laboratory-scale district heating network. The verification shows an average normalized root mean square error of 0.39 in the mass flow rates delivered to the buildings, and 0.15 in the network return temperature. Furthermore, the ability of the proposed modeling technique to rapidly generate models characterizing different network configurations is demonstrated through its application to topology optimization. The optimal design, obtained via a branch and bound algorithm, reduces network heat losses by 15\% percent as compared to the conventional length-minimized topology. 
\end{abstract}
\maketitle

\section{Introduction}

District heating networks (DHNs) offer numerous advantages over traditional heating methods, including reduced carbon emissions, flexible integration of a variety of intermittent renewable energy sources, and economies of scale, reducing the cost of heating buildings in urban environments. They are being increasingly adopted in cities of all sizes to increase energy efficiency and transition away from less sustainable heat sources. However, the performance of DHNs can be greatly improved by the implementation of advanced control strategies which consider individual users demands to deliver heat more effectively throughout the network \cite{delorenziSetupTestingSmart2020}. The first step to design such control strategies is the development of a control-oriented model that considers demand at the individual building level while being implementable on large-scale networks with the limited network data available.\par %While a lot of research has been done on modeling DHNs \cite{allegriniReviewModellingApproaches2015}, no existing modeling techniques have these desired features.\par
One existing approach is to use data-driven techniques to simulate the behavior of the system. For example, machine learning has been used to forecast the heat demands placed on a district heating network \cite{potocnikMachinelearningbasedMultistepHeat2021}. While shown to provide an accurate 48 hour prediction window, this model does not provide data on individual buildings demands and requires 3 years of data to train the model. Additionally, black box techniques have been used to learn acceptable fluid flow behaviors to enable the diagnosis of leakage faults that occur in the network \cite{xueMachineLearningbasedLeakage2020}. In one model, the steady-state heat losses in the piping network are estimated using a genetic algorithm trained using data from smart metering devices located throughout the network \cite{wangNewModelOnsite2018}. However, in most DHNs, a very limited amount of granular data is available on the dynamics within the network. Therefore, a black box modeling approach will not be effective for designing real-time control for the individual users throughout the network, and a physics-based approach is preferable.\par
Alternatively, a physics-based approach can be used where the behavior of all relevant components in the network are characterized thought first-principle equations. For example, computational fluid dynamics tools have been used to model the heat delays experienced by users different distances from the heat source \cite{zhaoInfluencingParametersAnalysis2019}. A modified finite volume technique was also developed, discretizing each pipe into segments \cite{betancourtschwarzModifiedFiniteVolumes2019}. A detailed pipe-oriented model has been developed in Modelica, where the time delay, spacial temperature distribution, advection, and thermal inertia are all considered in the pipe model \cite{vanderheijdeDynamicEquationbasedThermohydraulic2017}. While these models are comprehensive and can be used to accurately capture the DHN's behavior, calibrating and performing simulations is time-consuming, and would need to be repeated if the network configuration is modified. Moreover, high fidelity models are not very practical for control design, where the model's ability to scale to networks with hundreds of users is crucial.\par
Furthermore, some physics-based models have been derived targeting specific aspects of the network control. For example, one model aggregates regions of the network into substations where the behavior of multiple users are considered as a group, allowing for the fast simulation of large networks \cite{salettiControlorientedScalableModel2022}. However, the granularity provided by this model is not sufficient for the control of individual buildings. A substation-based model was also used to determine the level of thermal flexibility available in the network, allowing for the shifting of loads based on available heat sources \cite{vandermeulenSimulationbasedEvaluationSubstation2020}. Modeling for demand-side control has also been considered, where the heating demand of the entire network has been gathered into a single lumped parameter based on meteorological observations \cite{feltenIntegratedModelCoupled2020}. A model of the flexible heat capacity of a DHN has been developed in order to optimize the ability to better utilize renewable, intermittent heat sources \cite{huoOperationOptimizationDistrict2022}.\par
Another well-explored modeling approach for DHNs is a graph-based approach. Due to the inherent structure of a DHN, with long stretches of pipe connecting individual user nodes, many sources have chosen to represent the topology of a DHN as a graph. For example, a graph-based approach has been used for model order reduction, where spectral clustering of the network graph was used to group users into well connected subgroups \cite{simonssonReducedOrderModelingThermal2022}. A graph-based model has also been used to reduce the complexity of the fluid-flow calculation, mixing a black-box and physics-based approach \cite{guelpaCompactPhysicalModel2019}. Graph based approaches have also been used in the design of DHN configurations to minimize installation costs \cite{wackEconomicTopologyOptimization2023} and pressure drops \cite{suOptimizingPipeNetwork2022}.\par
The flexibility and scalability of graph-based modeling techniques have also allowed them successfully employed in a variety of other control applications. For example, a graph-based approach was effectively used to develop a control-oriented model of the various components in an aircraft \cite{williamsDynamicalGraphModels2017}. Graph-based models have also been used to partition large-scale systems into subsystems for use in distributed model predictive control \cite{jogwarDistributedControlArchitecture2019}.\par
In summary, while a variety of modeling approaches have been developed, none are designed to address building-level demand while remaining feasible for implementation on large-scale networks with the limited data available on granular network operation. The model equations presented in Saletti et al. \cite{salettiDevelopmentAnalysisApplication2020} provides an appropriate level of detail for control development. However, without a framework for interconnecting the model equations, developing the model of a single DHN is time-consuming and thereby considering a variety of configurations is challenging. Additionally, no validation is offered on the models ability to accurately capture the temperature dynamics in the network.\par
The main contribution of this paper is a technique for generating a state-space model of the network from a graph-based topology model of the network, with an algebraic calculation of the flow rates throughout the network. Temperature dynamics will be considered through a similar first-principles approach, with each pipe represented by a single energy conservation equation. The ability of the proposed methodology to capture the system dynamics is validated on data collected from a two-user lab-scale DHN. The automated nature of the proposed control-oriented modeling technique is demonstrated by considering a static design optimization problem, where a layout that minimizes the total length of the network is compared to one that considers the enthalpy losses during peak heat demands. The optimization problem will be solved by combining a graph theory problem formulation with the developed modeling technique, representing the problem as an optimum branch problem with prize collection constraints. It is solved using a branch and bound approach, where the proposed modeling technique will serve to calculate the enthalpy losses in each potential network configuration. \par
The remainder of the paper is organized as follows. \Cref{sec:model} presents an overview of the system and introduces the fundamental equations for the component models. In \cref{sec:network}, the graph-based technique is presented. The validation of the methodology and the experimental campaign are described in \cref{sec:validation}. The formulation of the energy minimization design problem is presented in \cref{sec:optimization}, while the results for a selected case study based on a real-world DHN are presented in \cref{sec:results}. Finally, conclusions are presented in \cref{sec:conclusion}.

\section{Modeling of Network Components}
\label{sec:model}
\begin{figure}
    \centering
    \includegraphics[width = 2.75in]{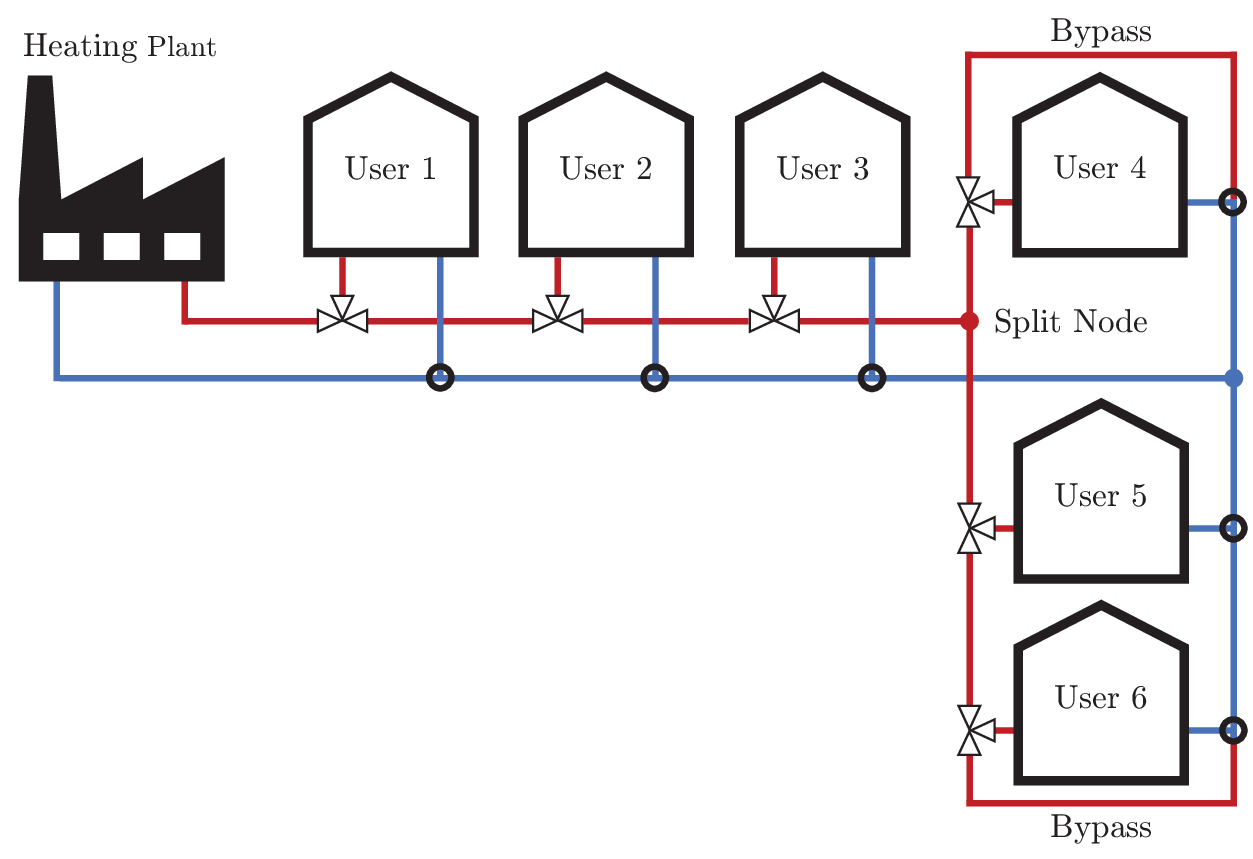}
    \caption{Sample six user DHN.}
    \label{fig:ex}
\end{figure}
DHNs consist of three main elements: the heating plant, the distribution network, and the users \cite{nussbaumerHandbookPlanningDistrict2020}. An example of a six user network is shown in \cref{fig:ex}. The water for the network is heated at the centralized heating plant which includes a pump to supply water through the distribution network to the users. The distribution network consists of two sets of pipes buried underground, the feeding and return networks. The feeding network delivers the heated water to the users and return network sends the cooled water back to the plant to be reheated. The feeding network is composed of pipe segments, that are connected at split nodes, where the water supply is divided between multiple branches of the network. The return network has mixing nodes to rejoin the flow from these branches to be returned to the plant.\par
\begin{figure}
\begin{subfigure}{.23\textwidth}
\vbox{\vspace*{1.7em}%
\centering{\includegraphics[width=\textwidth]{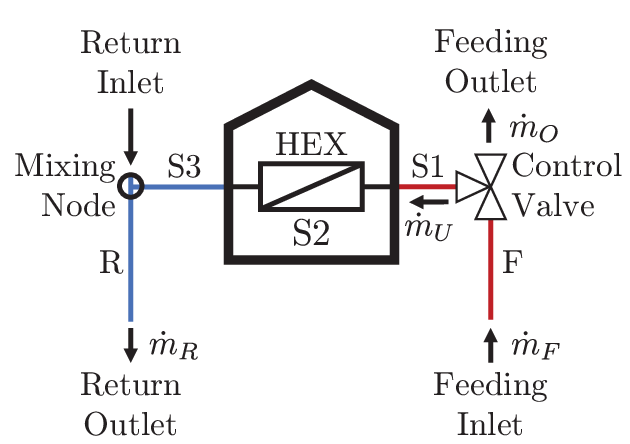}}%
\vspace*{1.7em}}%
\subcaption{User Only\label{fig:1Usera}}
\end{subfigure}%
\begin{subfigure}{.23\textwidth}
\vbox{\vspace*{1.7em}%
\centering{\includegraphics[width=\textwidth]{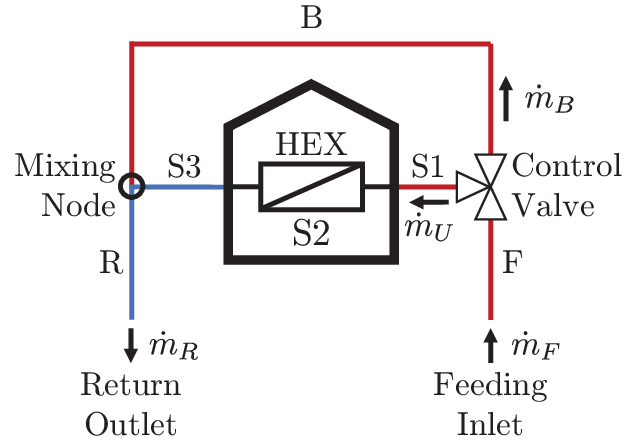}}%
\vspace*{1.7em}}%
\subcaption{User with bypass\label{fig:1Userb}}
\end{subfigure}%
\caption{Components of a single user loop with and without bypass segment.}
\label{fig:1User}
\end{figure}

Once the water has reached the user, a subnetwork of pipes together with a control valve, shown in \cref{fig:1User}, deliver water to the building. The control valve modulates the flow from the feeding segment $F$ into segment $S1$, which transports water to the heat exchanger, $S2$. Here, heat is removed from the network to be used by the building. The cooled water is returned through $S3$ to a mixing node, connected to return segment $R$. For users at the end of network branches, the feeding line is connected directly to the return line through a bypass segment, $B$, allowing any unused flow to be recirculated.\par

\subsection{Component Models}
The pipe segments making up the the feeding and return network, the split and mixing nodes, and the segments in a user block will be modeled via a first-principles based approach. First-principle models of DHNs components consider two domains, the thermal domain and the fluids domain \cite{salettiDevelopmentAnalysisApplication2020}.
The thermal domain, the relevant dynamics of the system, are considered using time-varying conservation of energy. The fluids domain is characterized by static pressure losses and continuity equations; this approach is chosen due to the relatively fast pressure dynamics when compared to changes in the network temperature \cite{bohmSimpleModelsOperational2002}.\par 
The temperature dynamics in a network segment are described by 
\begin{equation}
    \label{eq:Tpipe}
    \frac{d}{dt}T=\frac{\dot{m}}{\rho V}\left(T_{in}-T\right)-\frac{1}{\rho c_pV}\dot{Q}
\end{equation}
where $T$ is the bulk temperature throughout the pipe, $T_{in}$ is the inlet water temperature, $V$ is the volume of the pipe, $\dot{m}$ is the mass flow rate,  $\rho$ is the density,  $c_p$ heat capacity of the working fluid, and $\dot{Q}$ is the heat loss to either the environment or building.\par
For pipes in the feeding and return network, heat is lost to the environment and $\dot{Q}$ is given by
\begin{equation}
    \label{eq:Qamb}
    \dot{Q} = hA_s(T-T_{amb})
\end{equation}
where $T_{amb}$ is the ambient temperature, and $hA_s$ is the total conductive heat transfer coefficient between pipe and ground. The energy equation is then written in the compact form
\begin{equation}
    \label{eq:Tpipec}
    \frac{d}{dt}T=c_1 T_{in} + c_2 T_{amb} + c_3 T
\end{equation}
where 
\begin{subequations}
\label{eq:c}
\begin{equation}
    {c_1} = \frac{\dot{m}}{\rho V}
\end{equation}
\begin{equation}
    {c_2} = \frac{hA_s}{\rho c_pV}
\end{equation}
\begin{equation}
    {c_3} = -(c_1+c_2)
\end{equation}
\end{subequations}
For segment $S2$, the heat demand from the building is an input of the model, hence $\dot{Q} = \dot{Q}_b$, and the compact energy equation becomes
\begin{equation}
     \label{eq:Thex}
     \frac{d}{dt}T_{S2}=c_1T_{in}-c_1T_{S2}+c_4\dot{Q}_b
\end{equation}
where 
\begin{equation}
    \label{eq:c4}
    {c_4} = -\frac{1}{\rho c_pV}
\end{equation}\par
The pressure loss in any pipe segment is calculated by 
\begin{equation}
\label{eq:dP}
    \Delta P = \frac{1}{2\rho A_c^2}\left(k+\lambda\frac{L}{D}\right)\dot{m}^2 = \zeta\dot{m}^2
\end{equation}
where $k$ is the concentrated pressure loss coefficient, which accounts for changes in geometry in the pipe, and $\lambda$ is the distributed pressure loss coefficient, which accounts for friction with the pipe wall and is  dependent on pipe material. These two coefficients can either be calibrated separately using material properties and pipe geometry, or combined into one equivalent parameter $\zeta$, which is calibrated using experimental pressure loss data. Finally, $L$, $D$, and $A_c$ represent the pipes length, diameter, and cross sectional area respectively.\par
The mass flow is split between the branches in the network to cause the pressure losses in each branch connected to the same split node to be equal. This split is calculated using a pressure balance equation:
\begin{equation}
\label{eq:Pbal}
\Delta P_{Br}^{\{i\}}\left(\dot{m}_{in}^{\{i\}}\right) = \Delta P_{Br}^{\{j\}}\left(\dot{m}_{in}^{\{j\}}\right)\quad \forall i,j=1\dots n
\end{equation}
where $\dot{m}_{in}$ is the mass flow into the branch and $n$ is the total number of branches leaving the split node. Additionally the flow at each split node obeys the conservation of mass:
 \begin{equation}
 \label{eq:com}
     \dot{m}_{in} = \sum_{i = 1}^n\dot{m}^{\{i\}}
 \end{equation}
 where $\dot{m}_{in}$ is the mass flow rate into the split node. \par
 Finally, the heating plant is modeled as a heat supply with mass flow rate $\dot{m}_0$ and a controllable temperature $T_0$.
\subsection{User Block}
The component models are combined to generate a model of a single building connected to the network. This single user model serves as a building block when constructing the entire system. Here, a single user has five pipe segments, the three segments within the user, and the feeding and return line connecting the user to the network as shown in \cref{fig:1Usera}. Following this, the state equation for a single user $\{i\}$ is given by
\begin{equation}
    \label{eq:User}
    \frac{d}{dt}
    \begin{bmatrix}
    T_{F}^{\{i\}}\\T_{S1}^{\{i\}}\\T_{S2}^{\{i\}}\\T_{S3}^{\{i\}}\\T_{R}^{\{i\}}
    \end{bmatrix} = A_{U}^{\{i\}}\begin{bmatrix}
    T_{F}^{\{i\}}\\T_{S1}^{\{i\}}\\T_{S2}^{\{i\}}\\T_{S3}^{\{i\}}\\T_{R}^{\{i\}}
    \end{bmatrix}
    +E_{U}^{\{i\}}\begin{bmatrix}T_{amb}\\\dot{Q}_b^{\{i\}}\end{bmatrix}
\end{equation}
where $T_{Sj}^{\{i\}},\ j=1\dots 3$ are the temperatures of each pipe segment in the user, $T_{F}^{\{i\}}$ is the feeding line temperature, $T_{R}^{\{i\}}$ is the return line temperature, $A_{U}^{\{i\}}\in\mathbb{R}^{5\times 5}$ is the user's state transition matrix and $E_{U}^{\{i\}}$ is the user's external disturbance matrix. The matrix $A_{U}^{\{i\}}$ is constructed based on the interconnection between the pipe segments
\begin{equation}
    \label{eq:userA}
    A_{U}^{\{i\}}=
    \begin{bmatrix}
    {c_3}_F^{\{i\}}&0&0&0&0\\
    {c_1}_{S1}^{\{i\}}&{c_3}_{S1}^{\{i\}}&0&0&0\\
    0&{c_1}_{S2}^{\{i\}}&{-c_1}_{S2}^{\{i\}}&0&0\\
    0&0&{c_1}_{S3}^{\{i\}}&{c_3}_{S3}^{\{i\}}&0\\
    0&0&0&\frac{\dot{m}_{U}^{\{i\}}}{\dot{m}_{R}^{\{i\}}} {c_1}_{R}^{\{i\}}&{c_3}_{R}^{\{i\}}\\
    \end{bmatrix}
\end{equation}
where the flow mixing at the inlet of the return node is accounted for by mass flow rate ratio seen in the first nonzero term of the final row. 
The matrix $E_{U}^{\{i\}}$ is constructed from two column vectors, 
\begin{equation}
    \label{eq:userE}
    E_{U}^{\{i\}} \in \mathbb{R}^{6\times 2}=
    \begin{bmatrix}
    e_1^{\{i\}}&e_2^{\{i\}}
    \end{bmatrix}
\end{equation}
where the first column describes the heat transfer with the environment:
\begin{equation}
        e_1^{\{i\}} = 
    \begin{bmatrix}
    {c_2}_{F}^{\{i\}}&
    {c_2}_{S1}^{\{i\}}&
    0&
    {c_2}_{S3}^{\{i\}}&
    {c_2}_{R}^{\{i\}}
    \end{bmatrix}^T
\end{equation}
and second accounts for the heat flux into the building:
\begin{equation}
    e_2^{\{i\}} = 
    \begin{bmatrix}
    0&
    0&
    {c_4}_{S2}^{\{i\}}&
    0&
    0
    \end{bmatrix}^T
\end{equation}\par
For users connected to a bypass segment, as shown in \cref{fig:1Userb}, the model is augmented to include the additional temperature
\begin{gather}
    \frac{d}{dt}
    \begin{bmatrix}
    \overline{T}_U^{\{i\}}\\T_B^{\{i\}}
    \end{bmatrix} = A_B^{\{i\}} \begin{bmatrix}
    \overline{T}_U^{\{i\}}\\T_B^{\{i\}}
    \end{bmatrix}
    +E_B^{\{i\}}\begin{bmatrix}T_{amb}\\\dot{Q}_b^{\{i\}}\end{bmatrix}\\
    \label{eq:userAB}
    A_{B}^{\{i\}}\in\mathbb{R}^{6\times6}=\begin{bmatrix}
        A_{U}^{\{i\}} & \begin{matrix} 0^{4\times1}\\ \frac{\dot{m}_{B}^{\{i\}}}{\dot{m}_{R}^{\{i\}}}{c_1}_{R}^{\{i\}} \end{matrix}\\
        \begin{matrix} {c_1}_B^{\{i\}}&0^{1\times4}\end{matrix} & {c_3}_B^{\{i\}}
    \end{bmatrix}\\
    \label{eq:userEB}
    E_{B}^{\{i\}} =\begin{bmatrix}
        e_3^{\{i\}}&e_4^{\{i\}}\\ 
    \end{bmatrix} =\begin{bmatrix}
        e_1^{\{i\}}&e_2^{\{i\}}\\
        {c_2}_{B}^{\{i\}} & 0
    \end{bmatrix} 
\end{gather}
where $T_B^{\{i\}}$ is the temperature in the bypass, and $\overline{T}_U^{\{i\}}$ are the states given in \cref{eq:User}. 
\section{Modeling of Network Configurations}
\label{sec:network}
In this section, the method for combining the component models and user blocks to generate the state-space representation of a complete network is presented.
\subsection{Two-User Model}
\label{sec:2user}
\begin{figure}
    \centering
    \includegraphics[width = 3in]{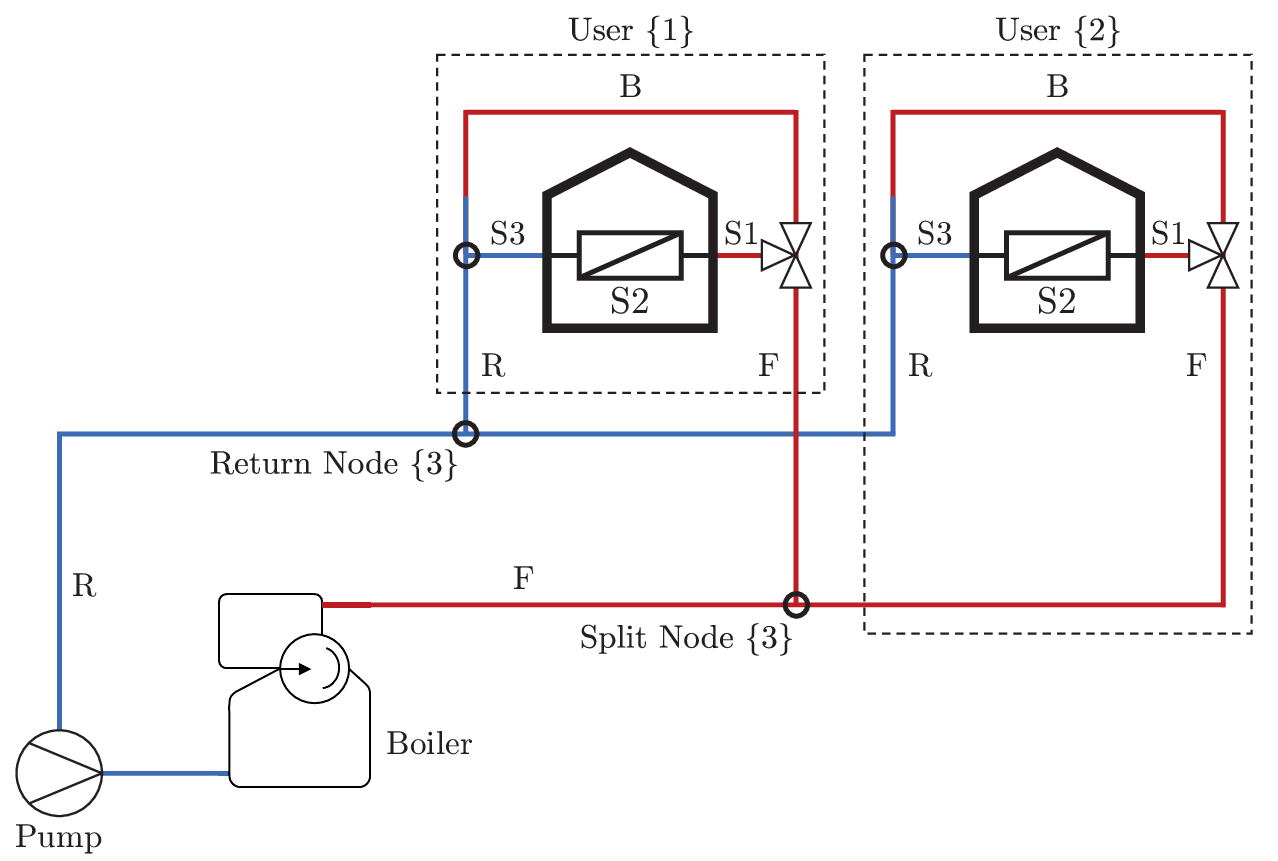}
    \caption{Layout of a two-user DHN with two parallel branches.}
    \label{fig:2User}
\end{figure}
An illustrative example is given where the two-user network shown in \cref{fig:2User} is considered. This network has two parallel branches meaning each user has its own bypass segment, and there is one split node. The temperature dynamics of this two-user network are given by
\begin{equation}
    \label{eq:2user}
    \frac{d}{dt} \begin{bmatrix}
    T_F^{\{3\}}\\ \overline{T}_{B}^{\{1\}}\\ \overline{T}_{B}^{\{2\}}\\T_R^{\{3\}}
    \end{bmatrix} = A_{aug}\begin{bmatrix}
    T_F^{\{3\}}\\ \overline{T}_{B}^{\{1\}}\\ \overline{T}_{B}^{\{2\}}\\T_R^{\{3\}}
    \end{bmatrix} +B_{aug}\begin{bmatrix}T_{0}\end{bmatrix} +E_{aug} \begin{bmatrix}T_{amb}\\\dot{Q}_b^{\{1\}}\\\dot{Q}_b^{\{2\}}\end{bmatrix}
\end{equation}
where $\overline{T}_{B}^{\{i\}},\ i=1,2$ is the state vector for each user with bypass as described by \cref{eq:User}, and $T_F^{\{3\}},T_R^{\{3\}}$ are the temperatures of the feeding and return lines connected to the split node. The corresponding state transition matrix is 
\begin{equation}
\label{eq:2userA}
    A_{aug} = \begin{bmatrix}
    {c_3}_F^{\{3\}} & 0^{1\times 6} & 0^{1\times 6} &0\\
    b_{21}^{\{1\}} & A_{B}^{\{1\}} & 0^{6\times 6} & 0^{6\times1} \\
    b_{21}^{\{2\}} & 0^{6\times 6}& A_B^{\{2\}} & 0^{6\times1} \\
    0 & b_{32}^{\{3,1\}} & b_{32}^{\{3,2\}} & {c_3}_R^{\{3\}}
    \end{bmatrix}
\end{equation}\par
As user blocks are connected to the overall network, additional vectors must be included in $A_{aug}$ to indicate these connections. The inlet temperature for a user's feeding line from the split node is given by $a_{21}^{\{i\}}$ for users without a bypass, and $b_{21}^{\{i\}}$ for users with a bypass segment, defined as
\begin{equation}
\label{eq:a21}
    a_{21}^{\{i\}} = \begin{bmatrix}
    {c_1}_{F}^{\{i\}}\\0^{4\times 1}\end{bmatrix},\  b_{21}^{\{i\}} =\begin{bmatrix}
        a_{21}^{\{i\}}\\0
    \end{bmatrix}\\
\end{equation}
The connection of the users return segment to the return network is give by $a_{32}^{\{i,j\}}$ for users without bypass segments and $b_{32}^{\{i,j\}}$ for users with a bypass segment, defined as
\begin{equation}
\label{eq:a32}
    a_{32}^{\{i,j\}} = \begin{bmatrix}
    0^{1\times 4}& \frac{\dot{m}_R^{\{j\}}}{\dot{m}_{R}^{\{i\}}}{c_1}_{R}^{\{i\}}
    \end{bmatrix},\ b_{32}^{\{i,j\}} =\begin{bmatrix}
        a_{32}^{\{i,j\}}&0 \end{bmatrix}
\end{equation}
where $j$ is the user number and $i$ is the return line to which the user is connected.\par 
The full network also includes the central heating plant, which supplies water with temperature, $T_0$, corresponding to the inlet temperature to the first feeding line in the network and is assumed to be controllable. The matrix $B_{aug}$ is given by
\begin{equation}
    \label{eq:2userB}
    B_{aug} = \begin{bmatrix}
    {c_1}_F^{\{3\}}\\0^{6\times 1}\\0^{6\times 1}\\0 
    \end{bmatrix}  
\end{equation}
The uncontrollable disturbances for the entire network are included in the $E_{aug}$ matrix as
\begin{equation}
    \label{eq:2userE}
    E_{aug} = \begin{bmatrix}
    {c_2}_F^{\{3\}}&0&0\\
    e_3^{\{1\}}&e_4^{\{1\}}&0^{6\times1}\\
    e_3^{\{2\}}&0^{6\times1}&e_4^{\{2\}}\\
    {c_2}_R^{\{3\}}&0&0
    \end{bmatrix}
\end{equation}
\subsection{Multi-User Model}
Larger network configurations are modeled following a similar procedure as demonstrated on the two-user example. In these cases, the challenge lies in efficiently converting connections between network elements into the state-space. To overcome this, the topology of the DHN configuration is represented by an unweighted rooted directed graph $ \mathcal{G} = (\mathcal{V},\mathcal{E})$. The dimensions of the network model are characterized by the number of users $n_u$ and split nodes $n_s$. The set of nodes $\mathcal{V} = \{v_0,\dots,v_{n_u+n_s}\}$ is decomposed into three sets $\mathcal{V} = \{v_0,\mathcal{V}_u,\mathcal{V}_s\}$, where $v_0$ is the node for the heating plant, $\mathcal{V}_n = \{v_1\dots,v_{n_u}\}$ contains the $n_u$ user nodes, and $\mathcal{V}_s = \{v_{n_u+1}\dots,v_{n_u+n_s}\}$ contains the $n_s$ split nodes. Additionally, let $\mathcal{V}_{leaf}\subset \mathcal{V}_u$ be the set of leaf nodes in the tree. All leaf users must have bypass segments, hence the total number of states in the network is
\begin{equation}
    n_p = 5n_u+n_s+\text{card}\{\mathcal{V}_{leaf}\}
\end{equation}
where $\text{card}\{\cdot\}$ is the cardinality of a set. The set of directed edges $\mathcal{E}=\{\varepsilon_1\dots \varepsilon_{n_u+n_s}\}$ represents the connections between the components in the network, based on the network's topology.

\begin{figure}
\begin{subfigure}{.23\textwidth}
\vbox{\vspace*{1.7em}%
\centering{\includegraphics[width=.5\textwidth]{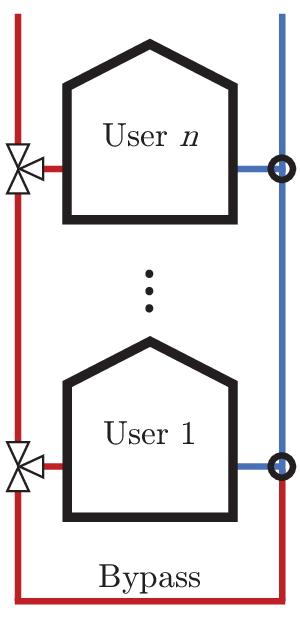}}%
\vspace*{1.7em}}%
\subcaption{Parallel configuration\label{fig:Brtypesa}}
\end{subfigure}%
\begin{subfigure}{.23\textwidth}
\vbox{\vspace*{1.7em}%
\centering{\includegraphics[width=.5\textwidth]{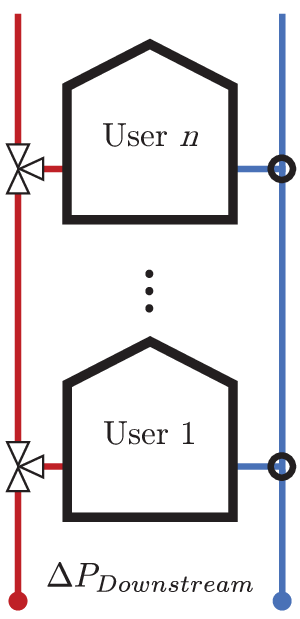}}%
\vspace*{1.7em}}%
\subcaption{Series configuration\label{fig:Brtypesb}}
\end{subfigure}%
\caption{Configurations of the two types of branches.}
\label{fig:Brtypes}
\end{figure}

The first step in calculating the network temperature dynamics is determining the mass flow rate in every pipe segment. Without loss of generality, this paper considers the case where the mass flow split at the split nodes has no actuation and the network has no booster pumps. The presence of booster pumps or additional actuators can be easily incorporated through additional conditions on the pressure balance equations. Additionally, it is assumed that the mass flow rate drawn by each user is known. This is a reasonable assumptions, as if this mass flow rate is unknown, it can be determined from the control valve position or heat demand. From the pressure losses in the pipe segments, an implicit system of equations is generated to find the mass flow rates in the network, while ensuring the pressure balance \cref{eq:Pbal} and conservation of mass \cref{eq:com} at each split node is maintained.\par
To facilitate this, groups of users connected directly to each other with no split nodes are identified as branches. Parallel branches (\cref{fig:Brtypesa}) are terminated by a user and are identified by the leaf nodes in the tree, while series branches (\cref{fig:Brtypesb}) end in split nodes and are characterized by the in-edge of said node. The pressure loss in any network branch of $n$ users is given by 
\begin{equation}
\label{eq:Pbr_gen}
    \Delta P_{Br(n)} = \sum_{i=0}^{n-1}2^i\zeta_F^{\{n-i\}}\left(\dot{m}_{in}-\sum_{j = 1}^i\dot{m}_U^{\{n+1-j\}}\right)^2+\phi
\end{equation}
where $\dot{m}_{in}$ is the flow rate into the branch, $\dot{m}_U^{\{i\}}$ is the flow drawn by user $i$. For a parallel branch, $\phi$ is given by
\begin{equation}
    \phi = 2^n \zeta_B\left(\dot{m}_{in}-\sum_{i = 1}^n\dot{m}_U^{\{i\}}\right)^2
\end{equation}
while for a series branch, $\phi$ is given by
\begin{equation}
    \phi = 2^n\zeta^F_{v_s}(\dot{m}_{out})^2 +2^n\sum_{i=1}^{n_{ds}} \Delta P^{Br}_{i}
\end{equation}
where $\dot{m}_{out}$ is the outlet flow, given by 
\begin{equation}
    \dot{m}_{out} = \dot{m}_{in}-\sum_{i=1}^n \dot{m}_U^{\{i\}}
\end{equation}
and $n_{ds} = \text{card}\{\mathcal{V}^-_{v_s}\}$ is the number of direct successors of the terminating split node $v_s$.
%\begin{proof}
This equation is constructed via induction. The pressure loss in any branch with $n$ users is given by
\begin{align}
    \Delta P_{Br(n)}=& \Delta P_F^{\{n\}} + \Delta P_U^{\{n\}}+\Delta P_{Br(n-1)}\\
    =& \Delta P_F^{\{n\}} + 2\Delta P_{Br(n-1)}
\end{align}
where $\Delta P_U^{\{n\}}=\Delta P_{Br(n-1)}$ due to the pressure balance at the control valve of user $n$.\par
The base case differs depending on branch type. As parallel branches include a bypass segment, the pressure loss for a single user in a parallel branch is given by
\begin{align}
    \Delta P_{Br(1)} =& \Delta P_F^{\{1\}} + \Delta P_U^{\{1\}}+\Delta P_{B}\\
    =& \Delta P_F^{\{1\}} + 2\Delta P_B\\
    =& \zeta_F^{\{1\}}(\dot{m}_{in}^{\{1\}})^2+2\zeta_B\left(\dot{m}_{in}^{\{1\}}-\dot{m}_U^{\{1\}}\right)^2
    \label{eq:P_p}
\end{align}
where the substitution $\Delta P_U^{\{1\}}=\Delta P_B$ is due to the pressure balance at the control valve.\par
However, because series branches have additional downstream branches, the pressure loss for a single-user series branch is given by 
\begin{align}
    \Delta P_{Br(1)}=& \Delta P_F^{\{1\}} + \Delta P_U^{\{1\}}+\Delta P_{F}^{\{v_s\}}+\sum_{i=1}^{n_{ds}} \Delta P_{Br}^{\{i\}}\\
    =& \Delta P_F^{\{1\}}+2\Delta P_F^{\{v_s\}} +2\sum_{i=1}^{n_{ds}} \Delta P_{Br}^{\{i\}}\\
     =& \zeta_F^{\{1\}}(\dot{m}_{in}^{\{1\}})^2 + 2\zeta_F^{\{v_s\}}(\dot{m}_{out})^2+2\sum_{i=1}^{n_{ds}} \Delta P_{Br}^{\{i\}}
     \label{eq:P_s}
\end{align}
where the substitution $\Delta P_U^{\{1\}}=\Delta P_{F}^{\{v_s\}}+\sum_{i=1}^{n_{ds}} \Delta P_{Br}^{\{i\}}$ is also due to the pressure balance at the control valve.
Combining \cref{eq:P_p,eq:P_s} with the induction term gives the final pressure equation as a function of branch type and number of users. 
%\end{proof}
Using \cref{eq:Pbr_gen} to describe the pressure loss in each network branch, the full pressure balance is resolved. The first step is to establish variables for the percent mass flow in each branch leaving the split nodes, $\alpha^{\{i\}}_{j},\forall\ i \text{ s.t. } v_i\in\mathcal{V}_s,\ j = 1\dots \text{card}\{\mathcal{V}^-_{v_i}\}$. Subsequently, each split node adds one conservation of mass equations given by
\begin{equation}
    \sum_{j=1}^{\text{card}\{\mathcal{V}^-_{v_i}\}}\alpha^{\{i\}}_j=1
\end{equation}
Additionally, each split node adds $\left(\text{card}\{\mathcal{V}^-_{v_i}\}-1\right)$ pressure balance equations:
\begin{equation}
    \Delta P_{Br}^{\{1\}} = \Delta P_{Br}^{\{j\}}\quad j=2\dots \text{card}\{\mathcal{V}^-_{v_i}\}
\end{equation}
where the pressure losses in each branch are found using \cref{eq:Pbr_gen} where $\dot{m}_{in}$ is the product of the upstream $\alpha$ values and the supply mass flow rate, $\dot{m}_0$. Hence, each split node adds $\text{card}\{\mathcal{V}^-_{v_i}\}$ unknowns and $\text{card}\{\mathcal{V}^-_{v_i}\}$ constraints, meaning there will be exactly one solution, which is calculated using a numeric solver due to the implicit form of the final equation.

\subsubsection{Temperature Modeling}
Once the mass flow rate in each pipe segment is obtained as a function of the user's flow demands $\dot{m}_{U}^{\{i\}}$, the connections between the nodes in the graph, along with the types of nodes, are used to identity the relationships between the states in the network. For each node, the graph provides the set of direct predecessors, $\mathcal{V}^-_{v_i}$, and direct successors, $\mathcal{V}^+_{v_i}$, which are used to generate the state-space model of the network
\begin{gather}
\label{eq:AugSS}
    \frac{d}{dt}T_{aug}  = A_{aug}T_{aug} +B_{aug} \begin{bmatrix}T_0 \end{bmatrix} +E_{aug} \begin{bmatrix} T_{amb}\\\dot{Q}_b^{\{1\}}\\\vdots\\\dot{Q}_b^{\{n_u\}} \end{bmatrix}\\
\begin{multlined}
    T_{aug} = \left[\begin{matrix} T_F^{\{n_u+1\}}& \dots& T_F^{\{n_u+n_s\}},& \overline{T}_{U,B}^{\{1\}}&\dots\end{matrix}\right.\\
    \left.\begin{matrix}  \overline{T}_{U,B}^{\{n_u\}},& T_R^{\{n_u+1\}}& \dots& T_R^{\{n_u+n_s\}} \end{matrix}\right]^T
\end{multlined}
\end{gather}
The augmented state matrix $A_{aug}$, originally defined in \cref{eq:2userA} is generalized for a network of any size by
\begin{equation}
    \label{eq:fullA}
    A_{aug} = \begin{bmatrix}
    A_{11} & A_{12} & 0^{n_s\times n_s}\\
    A_{21} & A_{22} & A_{23}\\
    0^{n_s\times n_s} & A_{32} & A_{33}\end{bmatrix}
\end{equation}
and it has seven nonzero submatrices, each of which characterize a different set of interconnections in the generalized network:
\begin{enumerate}
    \item $A_{11}\in\mathbb{R}^{n_s\times n_s}$: split nodes to feeding network
    \item $A_{12}\in\mathbb{R}^{n_s\times n_p}$: split nodes' feeding lines to users
    \item $A_{21}\in\mathbb{R}^{n_p\times n_s}$: users' feeding lines to split nodes
    \item $A_{22}\in\mathbb{R}^{n_p\times n_p}$: user blocks and users to users
    \item $A_{23}\in\mathbb{R}^{n_p\times n_s}$ split nodes' return line to users
    \item $A_{32}\in\mathbb{R}^{n_s\times n_p}$ users' return lines to split nodes
    \item $A_{33}\in\mathbb{R}^{n_s\times n_s}$ split nodes to the return network
\end{enumerate}\par
Complex network configurations have additional connection types beyond those seen in the two-user case, and populating the system matrices requires defining additional vectors to describe these interconnections. Specifically, $A_{21}$ contains terms $a_{12}^{\{i\}}$ defined by 
\begin{equation}
    a_{21}^{\{i\}} =  \begin{bmatrix}
    {c_1}_{F}^{\{i\}}&0^{1\times 4}\end{bmatrix}
\end{equation}
$A_{22}$ contains terms $a_{R22}^{\{i,j\}},\ b_{R22}^{\{i,j\}}$ defined by
\begin{equation}
    a_{R22}^{\{i,j\}} = \begin{bmatrix}
        0^{4\times4} & 0^{4\times1}\\
        0^{1\times4} & \frac{\dot{m}_R^{\{j\}}}{\dot{m}_R^{\{i\}}}{c_1}_R^{\{i\}}
    \end{bmatrix},\ 
    b_{R22}^{\{i,j\}} = \begin{bmatrix}
        a_{R22}^{\{i,j\}} & 0^{5\times1}
    \end{bmatrix}
\end{equation}
and terms $a_{F22}^{\{v_i\}},\ b_{F22}^{\{v_i\}}$ defined by
\begin{equation}
    a_{F22}^{\{i\}} = \begin{bmatrix}
        {c_1}_F^{\{i\}} & 0^{1\times4}\\
        0^{4\times1} & 0^{4\times4}
    \end{bmatrix},\ b_{F22}^{\{i\}} = \begin{bmatrix}
        a_{F22}^{\{i\}}\\
        0^{1\times5}
    \end{bmatrix}
\end{equation}
$A_{23}$ contains terms $a_{23}^{\{i,j\}}$, defined by
\begin{equation}
    a_{23}^{\{i,j\}} =  \begin{bmatrix}
    \frac{\dot{m}_R^{\{j\}}}{\dot{m}_R^{\{i\}}}{c_1}_{R}^{\{i\}}\\0^{4\times 1}\end{bmatrix}
\end{equation}
The detailed algorithm for generating the components of the $A_{aug}$ matrix is given in \cref{alg:A}.\par
The $B_{aug}$ and $E_{aug}$ matrices presented in \cref{eq:2userB} and \cref{eq:2userE} respectively are similarly generalized for a network of any size. The $B_{aug}$ matrix is supplemented to contain terms for $v_i\in\mathcal{V}^+_{v_0}$ and is found according to \cref{alg:B}. The $E_{aug}$ includes terms for the heat losses of every pipe in the network and includes heat transferred into every building served by the network. The methodology for generating $E_{aug}$ is presented in \cref{alg:E}.
\begin{algorithm}
    \caption{Generation of $A_{aug}$ submatrices}
    \label{alg:A}
    \begin{algorithmic}[1]
    \ForAll{$i$ s.t. $v_i \in \mathcal{V}_u\backslash\mathcal{V}_{leaf}$}
        \State{$A_{22}(i,i) \gets A_U^{\{i\}}$}
        \State{$A_{22}(i,j) \gets a_{R22}^{\{i,j\}}\ \forall\ j$ s.t. $v_j\in\mathcal{V}_{v_i}^+\cap\mathcal{V}_u\backslash\mathcal{V}_{leaf}$}
        \State{$A_{22}(i,j) \gets b_{R22}^{\{i,j\}}\ \forall\ j$ s.t. $v_j\in\mathcal{V}_{v_i}^+\cap\mathcal{V}_{leaf}$}
        \State{$A_{23}(i,j-n_u) \gets a_{23}^{\{i,j\}}\ \forall\ j$ s.t. $v_j\in\mathcal{V}_{v_i}^+\cap\mathcal{V}_s$}
        \State{$A_{21}(i,j-n_u) \gets a_{21}^{\{i\}}\ \forall\ j$ s.t. $v_j\in\mathcal{V}_{v_i}^-\cap\mathcal{V}_s$}
        \State{$A_{22}(i,j) \gets a_{F22}^{\{i\}}\ \forall\ j$ s.t. $v_j\in\mathcal{V}_{v_i}^-\cap\mathcal{V}_u$}
    \EndFor
    \ForAll{$i$ s.t. $v_i\in\mathcal{V}_{leaf}$}
        \State{$A_{22}(i,i) \gets A_B^{\{i\}}$}
        \State{$A_{21}(i,j-n_u) \gets b_{21}^{\{i\}}\ \forall\ j$ s.t. $v_j\in\mathcal{V}_{v_i}^-\cap\mathcal{V}_s$}
        \State{$A_{22}(i,j) \gets b_{F22}^{\{i\}}\ \forall\ j$ s.t. $v_j\in\mathcal{V}_{v_i}^-\cap\mathcal{V}_u$}
    \EndFor
    \ForAll{$i\ \textbf{s.t.}\ v_i \in \mathcal{V}_s$}
        \State{$A_{11}(i-n_u,i-n_u) \gets {c_3}_F^{\{i\}}$}
        \State{$A_{33}(i-n_u,i-n_u) \gets {c_3}_R^{\{i\}}$}
        \State{$A_{11}(i-n_u,j-n_u) \gets {c_1}_F^{\{i\}}\ \forall\ j$ s.t. $v_j\in\mathcal{V}_{v_i}^-\cap\mathcal{V}_s$}
        \State{$A_{12}(i-n_u,j) \gets a_{12}^{\{i\}}\ \forall\ j$ s.t. $v_j\in\mathcal{V}_{v_i}^-\cap\mathcal{V}_u$}
        \State{$A_{32}(i-n_u,j) \gets a_{32}^{\{i,j\}}\ \forall\ j$ s.t. $v_j\in\mathcal{V}_{v_i}^+\cap\mathcal{V}_u\backslash\mathcal{V}_{leaf}$}
        \State{$A_{32}(i-n_u,j) \gets b_{32}^{\{i,j\}}\ \forall\ j$ s.t. $v_j\in\mathcal{V}_{v_i}^+\cap\mathcal{V}_{leaf}$}
        \State{$A_{33}(i-n_u,j-n_u) \gets \frac{\dot{m}_R^{\{j\}}}{\dot{m}_R^{\{i\}}}{c_1}_R^{\{i\}}\ \forall\ j$ s.t. $v_j\in\mathcal{V}_{v_i}^+\cap\mathcal{V}_s$}
    \EndFor
    \end{algorithmic}
\end{algorithm}

\begin{algorithm}
    \caption{Generation of $B_{aug}$}
    \label{alg:B}
    \begin{algorithmic}[1]
    \State{$B_{aug}\in\mathbb{R}^{(2n_s+n_p)\times 1}$}
    \ForAll{$i$ s.t. $v_i\in\mathcal{V}^-_{v_0}$}
        \If{$i\in\mathcal{V}_u\backslash\mathcal{V}_{leaf}$}
            \State{$B_{aug}(i+n_s,1)\gets\begin{bmatrix}
                {c_1}_F^{\{i\}}\\ 0^{4\times1}
            \end{bmatrix}$}
        \ElsIf{$i\in\mathcal{V}_{leaf}$}
        \State{$B_{aug}(i+n_s,1)\gets\begin{bmatrix}
                {c_1}_F^{\{i\}}\\ 0^{5\times1}
            \end{bmatrix}$}
        \ElsIf{$i\in\mathcal{V}_{s}$}
            \State{$B_{aug}(i-n_u,1)\gets {c_1}_F^{\{i\}}$}
        \EndIf
    \EndFor
    \end{algorithmic}
\end{algorithm}

\begin{algorithm}
    \caption{Generation of $E_{aug}$}
    \label{alg:E}
    \begin{algorithmic}[1]
    \State{$E_{aug}\in\mathbb{R}^{(2n_s+n_p)\times (1+n_u)}$}
    \ForAll{$i$ s.t. $v_i\in\mathcal{V}_{u}\backslash\mathcal{V}_{leaf}$}
        \State{$E_{aug}(i+n_s,1)\gets e_1^{\{i\}}$}
        \State{$E_{aug}(i+n_s,i+1)\gets e_2^{\{i\}}$}
    \EndFor
    \ForAll{$i$ s.t. $v_i\in\mathcal{V}_{leaf}$}
        \State{$E_{aug}(i+n_s,1)\gets e_3^{\{i\}}$}
        \State{$E_{aug}(i+n_s,i+1)\gets e_4^{\{i\}}$}
    \EndFor
    \ForAll{$i$ s.t. $v_i\in\mathcal{V}_{s}$}
        \State{$E_{aug}(i-n_u,1)\gets {c_2}_F^{\{i\}}$}
        \State{$E_{aug}(i+n_s,1)\gets {c_2}_R^{\{i\}}$}
    \EndFor
    \end{algorithmic}
\end{algorithm}
\section{Validation of the Dynamic Model}
\label{sec:validation}
Due to the lack of available high spatial and temporal resolution data about the dynamic operation of full-scale DHNs, a lab-scale experimental DHN was developed by the authors and presented in \cite{blizardDynamicallySimilarLabscale2023}. The lab-scale DHN was designed via the Buckingham $\pi$ theorem to match the dynamic response of a full-scale DHN while performing accelerated tests. The laboratory setup allows for the collection of detailed temperature and flow data used in this paper to validate the ability of this modeling approach to capture the temperature response of a real-world DHN. The developed test bench has the same configuration as two-user example provided in \cref{sec:2user}.\par
\begin{figure}
    \centering
    \includegraphics[width = 3.33in]{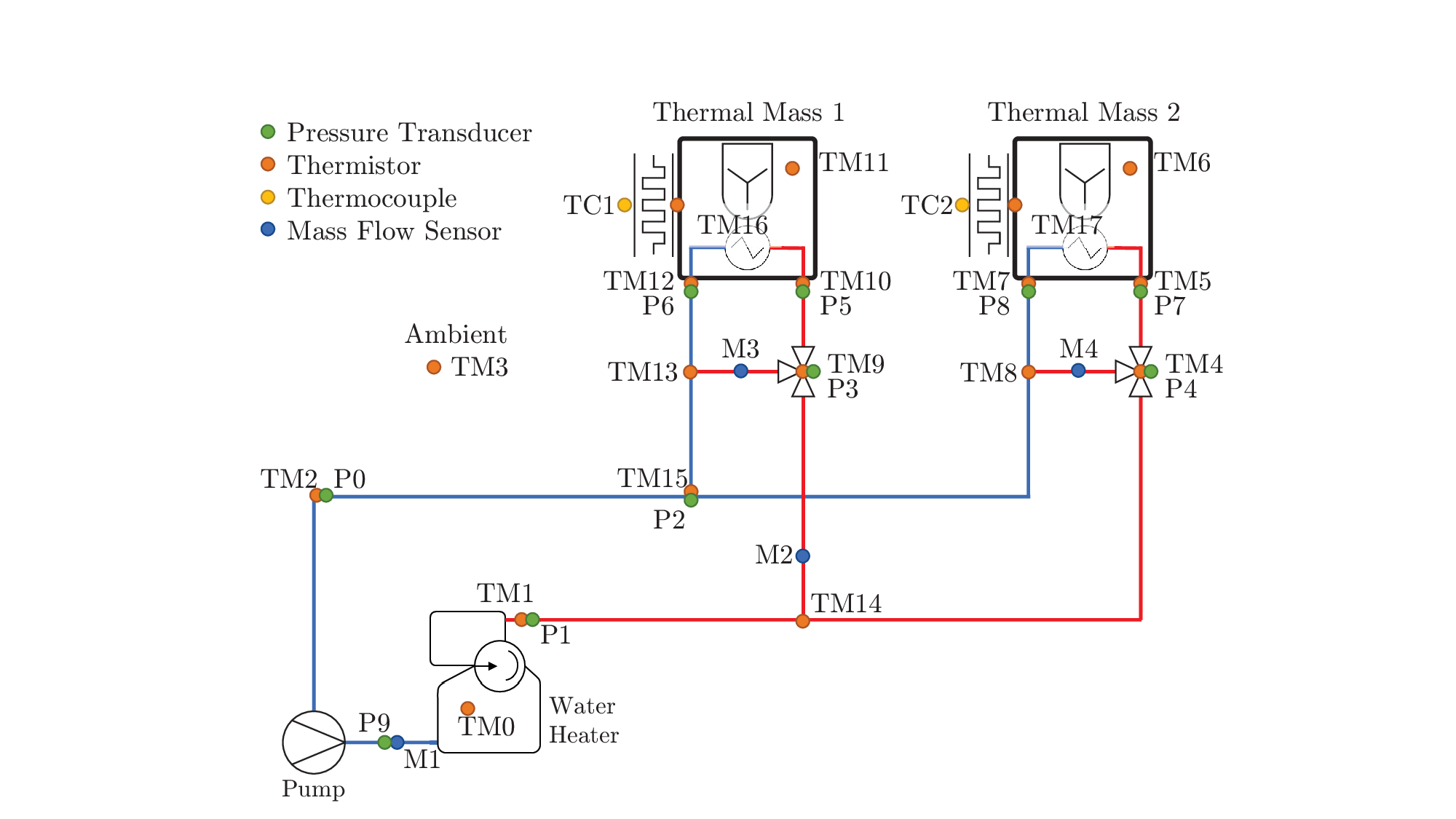}
    \caption{Diagram of the two-user lab-scale DHN with sensor locations labeled.}
    \label{fig:sensor}
\end{figure}

Data from this network is collected using the 17 thermistors, 9 pressure transducers, and 4 mass flow rate sensors located at key points in the network, as shown in \cref{fig:sensor}. The data was collected using two synchronized data acquisitions systems, with sampling rates of 1 Hz. Furthermore, the lab-scale DHN is equipped with two characterized control valves controlled through PID controllers, which regulate the flow to the thermal masses to achieve the desired thermal mass temperature setpoints. \par
A similar PID controller is implemented in the simulation, where the difference between the simulated building temperature and experimental building temperature is used to set the simulated mass flow rate delivered to the users. The simulation is performed in discrete time, using the bilinear transform to discretize the state space model, with a time step of one second.\par
To simplify the integration of the heat losses in the simulation, the temperatures of the buildings are appended to the developed state-space model. The temperature of a single building is given by
\begin{equation}
   \frac{d}{dt}T_{b}  = \frac{\left(hA_s\right)_{S2}}{\left(\rho c_pV\right)_{b}}\left(T_{S2}-T_b\right) - \frac{\left(hA_s\right)_b}{\left(\rho c_pV\right)_b}\left(T_b-T_{amb}\right)
\end{equation}
From this equation, the heat demand of the building can be calculated according to
\begin{equation}
    \dot{Q}_b = \left(hA_s\right)_{S2}\left(T_{S2}-T_{b}\right)
\end{equation}
where $\left(hA_s\right)_{S2}$ is the convective heat transfer coefficient of the heat exchanger and $T_{b}$ is the building temperature.\par
\begin{figure}
    \centering
    \includegraphics[width = 3in]{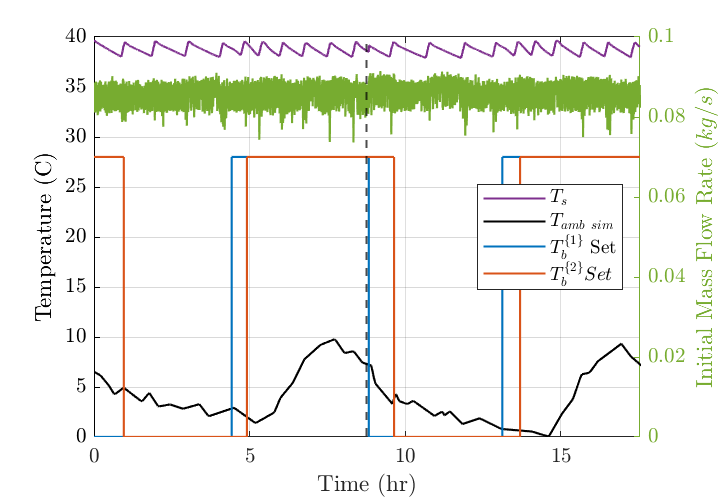}
    \caption{Profile of conducted experiment}
    \label{fig:exp}
\end{figure}

 
The data used to calibrate and validate the model was collected from 17.5 hour experiment, based on the one used in \cite{salettiDevelopmentAnalysisApplication2020}, which is equivalent to two days in a full-scale system. Realistic occupancy cycles were used to provide the temperature set points for the building, where the buildings are allowed to cooled overnight when unoccupied. This setpoint profile allows a variety of conditions to be explored, including heating, cooling, and temperature sustaining operation of a DHN. The simulation was performed using a realistic ambient temperature profile, representative of the average winter temperatures in northern Italy. The temperature set-points, ambient temperature profile, and supply temperature, are presented in \cref{fig:exp}. Additionally, the mass flow rate supplied to the network $\dot{m}_0$ is also presented. The collected data was split into two parts, indicated by the dashed line. The first data set was used to calibrate the model parameters while the second set was used to validate the model’s performance. \par
The calibration data was used to select the heat transfer coefficients $h$ for each pipe segment that minimize the sum of the root mean square error between the actual temperatures collected by the 16 thermistors and the simulated temperatures. In the calibration, the mass flow rate split between the network branches was taken from the data collected by M2. The resulting $h$ values for each pipe segments are provided in \cref{tbl:h}, and the normalized root mean square errors (nRMSE), normalized with respect to the value's range, for a selected subset of the sensors are provided in \cref{tbl:nRMSE}. \par
\begin{table}
\caption{Calibrated $h$ values ($[W/K]$).}
\label{tbl:h}
\centering
\begin{tabular}{ccc}
\toprule
Node&Segment&$h\ [W/K]$\\
\midrule
\multirow{6}{*}{$\{1\}$}&F&0.0102\\
&S1&301.1\\
&S2&$1.627\times10^3$\\
&S3&0.0163\\
&R&0.0134\\
&B&0.0036\\
\midrule
\multirow{6}{*}{$\{2\}$}&F&0.0040\\
&S1&244.0\\
&S2&$1.408\times10^3$\\
&S3&0.0126\\
&R&0.0121\\
&B&0.0036\\
\midrule
\multirow{2}{*}{$\{3\}$}&F&3.759\\
&R&0.0016\\
\bottomrule
\end{tabular}
\end{table}

\begin{figure}
    \centering
    \includegraphics[width = 3in]{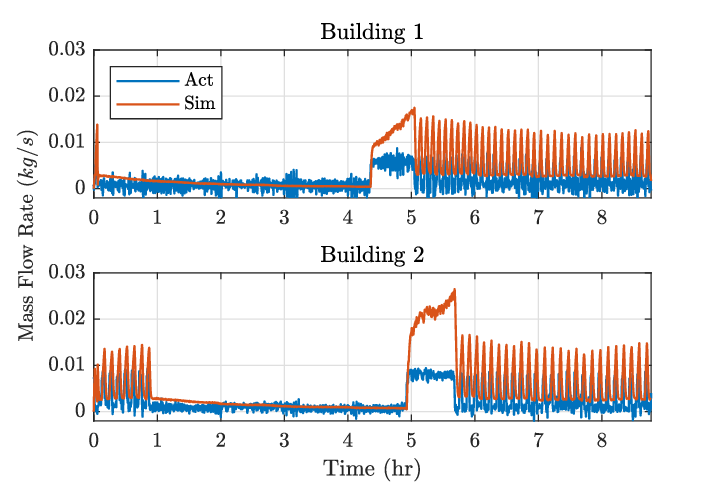}
    \caption{Comparison of the mass flow rate through the heat exchangers}
    \label{fig:mdot}
\end{figure}

\begin{figure}
    \centering
    \includegraphics[width = 3in]{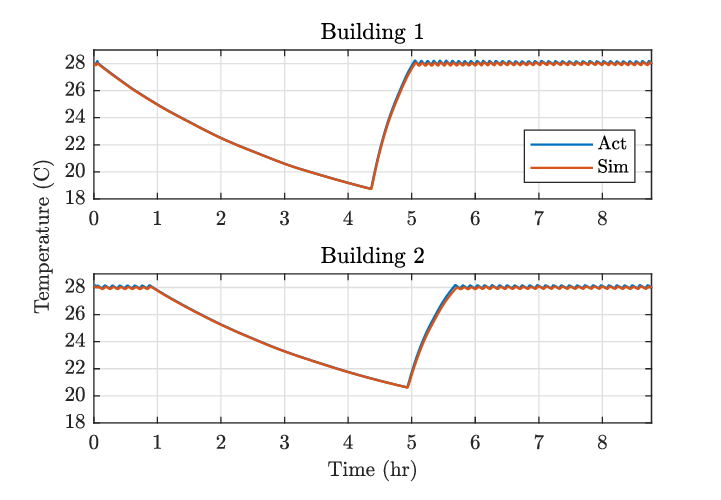}
    \caption{Comparison of the temperature response of the two buildings}
    \label{fig:ThM}
\end{figure}

\begin{figure}
     \centering
     \includegraphics[width = 3in]{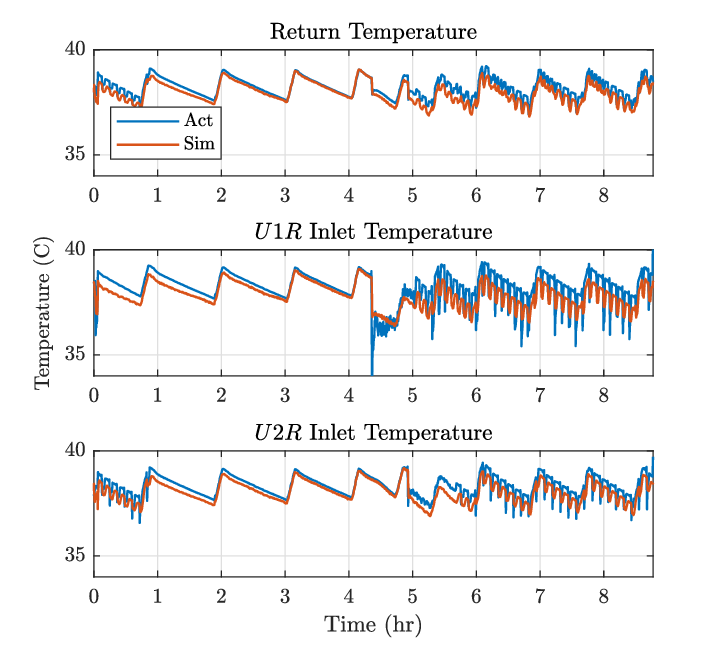}
     \caption{Comparison of temperatures throughout the network}
     \label{fig:pipes}
 \end{figure}

\begin{table}
\caption{nRMSE of network parameters.}
\label{tbl:nRMSE}
\centering
\begin{tabular}{l c c}
\toprule
Value&Calibration nRMSE & Validation nRMSE\\
\midrule
$T_b^{\{1\}}$&0.0085&0.0085\\
$T_b^{\{2\}}$&0.0112&0.0121\\
$T_R^{\{3\}}$&0.1432&0.1498\\
$T_{R\ in}^{\{1\}}$&0.0979&0.0769\\
$T_{R\ in}^{\{2\}}$&0.1126&0.1110\\
$\dot{m}_U^{\{1\}}$&0.2144&0.4732\\
$\dot{m}_U^{\{2\}}$&0.1940&0.3144\\
\bottomrule
\end{tabular}
\end{table}
The calibrated $h$ values are then used in a closed loop validation as shown in \cref{fig:mdot,fig:ThM,fig:pipes}. In this simulation, the mass flow split between the branches is calculated from the pressure losses in the network. The mass flow rate supplied to the users is compared in \cref{fig:mdot}. The temperature response of the two buildings are compared in \cref{fig:ThM}, while the temperatures at the outlet of return network, and at the inlet of the user's return segments are compared in \cref{fig:pipes}. The nRMSE are summarized in \cref{tbl:nRMSE}. Additionally, the results from the other network pipe segments were compared and have similar dynamics and errors to those shown.\par
As shown in \cref{tbl:h}, the heat transfer coefficients of the various segments are different, even though the pipe segments has the same insulation and material properties. This indicates that the model’s accuracy is dependent on proper calibration of these parameters, and if the network configuration changes, or additional users are added, the calibration procedure will have to be repeated. Furthermore, this calibration only considers a winter profile, as this is where the most potential for energy savings exists. Additional calibrations can be performed for different ambient conditions.\par

\section{Solving the Energy-Reduced Design Problem}
\label{sec:optimization}
The model's performance in optimization problems is critical for its application to novel control design. Additionally, the flexibility of the modeling technique when considering a variety of different layouts will allow it to easily adapt to changes in network topology, such as during faults, or as the system is partitioned for distributed control. This capability is demonstrated by rapidly generate models of various network configurations needed to resolve an optimal design problem.
%The flexibility of the modeling technique when considering a variety of models of different layouts, and its use in optimization problems is also critical for its application to novel control design. This capability is demonstrated by using the modeling technique to calculate the enthalpy losses in various DHN configurations, with the goal of choosing the network configuration that minimize the energy lost in the supply network during peak heating demand and steady-state operation. 
\subsection{Problem Formulation}
\begin{figure}
    \centering
    \includegraphics[width=3.33in]{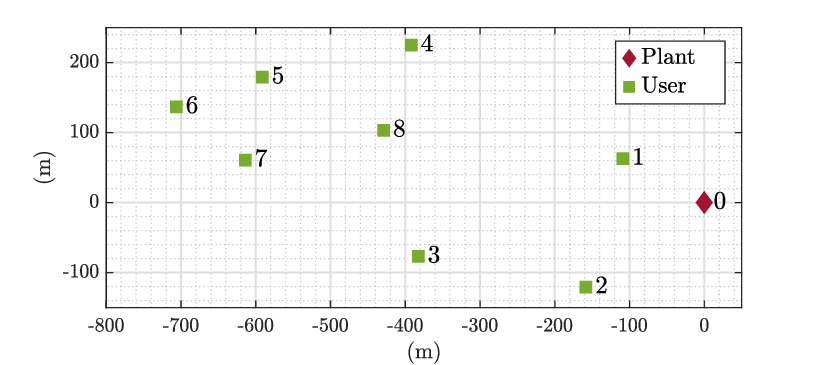}
    \caption{Map of 8 users to be connected, selected from the University of Parma DHN.}
    \label{fig:map}
\end{figure}

The goal of the optimal design problem is minimizing the enthalpy losses of the pipes in the supply network by changing the network layout, $\mathcal{G}$, where the change in enthalpy gives a measure of thermal power lost to the environment. The problem is formulated as
\begin{equation}
\label{eq:min}
\begin{gathered}
    C^*= \min_{\mathcal{G}}\sum_{\forall\ i\text{ s.t. } v_i\in\mathcal{V}} \left(\dot{m}c_p(T_{in}-T)\right)^{\{i\}} \\
     \text{ s.t. } v_0\in\mathcal{V}\\
     \text{card}\left(\mathcal{V}_u\right) = n_u
     \end{gathered}
\end{equation}
In this problem, the final network is assumed to be ideally pressure balanced, meaning the same mass flow rate will be delivered to each user. This is a realistic assumption and can be achieved through the use of pressure regulators and booster pumps. Therefore $\dot{m}^{\{i\}} = \frac{n_{u(ds)}}{n_u}\dot{m}_{0}$ where $n_{u(ds)}$ is the number of users downstream from node $v_i$, including node $v_i$. The network is assumed to be supplied by water with a constant supply temperature of 80 C, and the ambient temperature is assumed to be -5 C. For simplicity, only two pipe diameters are considered, a larger diameter $D_L$ for feeding lines connect to more than one user, and a smaller diameter $D_S$ for pipes connected to a single user. However, the pipe diameter has a large impact on the overall heat transfer in the network \cite{piroutiEnergyConsumptionEconomic2013}, and an extension to increase the granularity of the diameters is easily made in the recursive structure of the solution. All pipes are assumed to have the same heat transfer coefficient per unit surface area, $h$. A table of selected values for this simulation is presented in \cref{table:params}, and a map of the users to be connected is presented in \cref{fig:map}. The selected user locations are taken from a subset of the buildings served by the University of Parma DHN, providing a realistic case study of user location and spread.\par

\begin{table}
\caption{Parameters for energy loss minimization problem.}
\label{table:params}
\centering
\begin{tabular}{l l l c}
\toprule
Parameter & Symbol & Value & Source\\
\midrule
Initial mass flow rate & $\dot{m}_0$& 20 $kg/s$& \cite{salettiDevelopmentAnalysisApplication2020}\\
Density &$\rho$ & 971 $kg/m^3$ & \cite{cimbala2006fluid}\\
Specific heat capacity & $c_p$ & 4179 $J/kgK$& \cite{cimbala2006fluid}\\
Supply temperature & $T_0$ & 80 $C$ & \cite{gabrielaitieneModellingTemperatureDynamics2007}\\
Ambient temperature & $T_{amb}$ & -5 $C$ & \cite{anconaApplicationDifferentModeling2019}\\
Large pipe diameter & $D_L$ & 0.40 $m$ & \cite{masatinEvaluationFactorDistrict2016}\\
Small pipe diameter & $D_S$ & 0.15 $m$ & \cite{masatinEvaluationFactorDistrict2016}\\
Heat transfer coefficient & $h$ & 1.5 $W/m^2K$ & \cite{masatinEvaluationFactorDistrict2016}\\
\bottomrule
\end{tabular}
\end{table}
\begin{figure}
    \centering
    \includegraphics[width =3.33in]{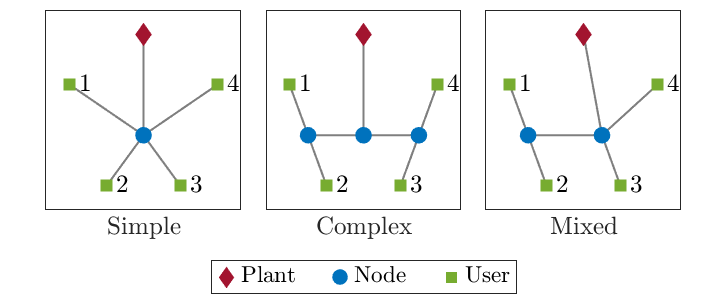}
    \caption{Sample of split node types for 4 user DHN}
    \label{fig:spltnds}
\end{figure}
Finally, to reduce the search space of potential layouts, assumptions must be made about how the users in network can be connected. In this paper, a series of candidate split nodes are generated a-priori to be used as potential split nodes in the final network graph. These split nodes are classified into three categories, simple, complex, and mixed. Simple split nodes connect to users directly and span the unique combinations of all subset of users in the network. Complex split nodes are nodes which connect other split nodes. These are generated recursively, keeping track of the users added in each addition to ensure no repetitions occur. Mixed split nodes are similar to complex split nodes, but contain direct connections to users. \Cref{fig:spltnds} shows an example of each type of split node, and the algorithm used to generate the candidate midpoints is presented in \cref{alg:spltnds}. In this problem, all split nodes are assumed to be located equidistant from the users or split nodes to which they connect. However, possible geographic limitations seen in the design of real-world heating networks can be considered by modifying the rules used to locate the candidate split nodes.

\begin{algorithm}
    \caption{Generation of Candidate Split Nodes}
    \label{alg:spltnds}
    \begin{algorithmic}[1]
    \State{\textbf{global} $\mathcal{V}_s,Pr_s$}
    \For{$i=1\dots n_u$}
        \State{Append $\mathcal{V}_{simple}$ to $\mathcal{V}_{s}$}
        \State{Append $v_i\in \mathcal{V}_{simple}$ to $Pr_{s}$}
    \EndFor
    \State{$i\gets2$}
    \While{$i<\text{card}\{\mathcal{V}_s\}$}
        \State{\Call{AddNode}{$v_i,Pr_{s}(i),i$}}
        \State{$i++$}
    \EndWhile

    \Procedure{AddNode}{$v,Pr,ii$}
        \If{$\text{card}\{Pr\}<n_u$}
        \State{$k\gets1$}
            \ForAll{$j<ii$ s.t. $Pr_s(j)\cap Pr=\varnothing$}
                \State{Create split node $v_{new}$ from $v$ and $\mathcal{V}_s(j)$}
                \State{Append $v_{new}$ to $\mathcal{V}_{s}$}
                \State{Append $Pr_s(j)\cup Pr$ to $Pr_{s}$}
                \State{\Call{AddNode}{$\begin{bmatrix} v & \mathcal{V}_s(j) \end{bmatrix},Pr_s(j)\cup Pr,ii+k$}}
                \State{$k++$}
            \EndFor
       \EndIf  
    \EndProcedure
    \end{algorithmic}
\end{algorithm}

\subsection{Solution Method}
The problem of connecting the users and candidate midpoints is considered as a dynamic optimum branching problem with a set of prizes and collection costs and is solved via a branch and bound algorithm. The goal of an optimum branching problem is to connect all nodes of a directed graph, the problem graph, in a minimum cost tree, and is a well-known problem in graph theory \cite{edmonds1967optimum}. In this formulation, the problem conditions are modified so that not all nodes must added to the final tree. Instead, the complete set of prizes, the unique set of $n_u$ users, must be collected \cite{manerbaTravelingPurchaserProblem2017}. Each node has a prize set associated with it, which denotes the subset of users connected to the node. The set of prize sets for all candidate split nodes is denoted as $Pr_s$. Additionally, some nodes also have a cost associated with them, so that when added to the tree, there is an additional cost beyond the edge cost to connect them. Finally, the problem is dynamic as the edge and node cost are variable based on how the tree is connected. The problem graph is formulated as follows. The nodes are the centralized plant, which will function as the root, the users, and candidate split nodes. The edges of the problem graph represent the  directed connection between all nodes and the root, and between all nodes that do not contain overlapping prize sets.\par
Due to the dynamic nature of the problem, a best-case scenario must be assumed in order to solve this problem using a branch and bound algorithm. This best-case scenario allows for a calculation of the lower bound cost (LBC) for the split nodes and the edges connecting the plant, split nodes, and users. Two factors which are only available after the generation of the complete graph influence the cost of the connection: the number of downstream users, and the inlet temperature supplied by the previous pipe. The number of downstream users affects the pipe diameter and the mass flow rate through the pipe. In both cases, a higher number of downstream users increases the enthalpy loss by increasing the necessary pipe diameter and mass flow rate. Therefore, the LBC assumes that no additional users are connected downstream. The other assumption is a bound on the maximum temperature drop in the network. Here an arbitrary maximum drop $\Delta T_{i}$ is assumed. With this, the LBC is calculated using $T_0-\Delta T_{i}$ as the inlet temperature. After the final minimum cost spanning tree is found, the maximum temperature drop $\Delta T_{max}$ is calculated, and the entire solution process is iterated if $\Delta T_i<\Delta T_{max}$. Using these bounding assumptions, two sets of costs are calculated, the LBC of connecting the users to the split nodes ($LBC_{s}$), which serve as the price for adding a split node to the tree, and the edge LBC  ($LBC_{\mathcal{E}}$), which provides a cost for adding a node to the tree.\par
After a complete minimum LBC tree is found, the true cost (TC) is calculated, using the modeling technique described in \cref{sec:model}. For this problem, because the goal is to minimize the heat lost to the environment during heat delivery, the return network is assumed to have the same characteristics of the feeding network, and the mass flow rate demanded by the users is assumed to be known, only the feeding lines must be considered. Therefore, the states for $S1, S2, S3, R$ are neglected leaving
\begin{gather}
    \frac{d}{dt}T_{red}  = A_{red} T_{red}+B_{red} \begin{bmatrix}T^{0} \end{bmatrix} +E_{red} \begin{bmatrix} T_{amb}\end{bmatrix}\\
    T_{red} = \begin{bmatrix} T_F^{\{n_u+1\}}&\dots&T_F^{\{n_u+n_s\}},&T_{F}^{\{1\}}&\dots&T_{F}^{\{n_u\}}\end{bmatrix}^T
\end{gather}
where $A_{red}$, $B_{red}$, and $E_{red}$ the augmented matrices where the rows and columns associated with the removed states are also removed. $A_{red}$ only contains $A_{11}$ and parts of the $A_{12}, A_{21}, A_{22}$ matrices. After a candidate tree is generated, the parameters $c_1, c_2, c_3$ are calculated using \cref{eq:c}. Then, because the problem considers steady state operation of the network, the dynamic term is set to zero, and the temperature equation is rearranged to give the steady state inlet and outlet temperatures using
\begin{equation}
\label{eq:Tred}
   T_{red} = -A_{red}^{-1}\left( B_{red} \begin{bmatrix}T_0 \end{bmatrix} +E_{red} \begin{bmatrix} T_{amb}\end{bmatrix} \right)
\end{equation}
where $A_{red}$ is invertible, as it can always be written as a lower-triangular matrix with all nonzero diagonal elements ${c_3}_{F}^{\{i\}},\ \forall i=1\dots n_s+n_u$ by reordering the nodes.
These temperatures are used to solve for the true cost of a particular network configuration.\par
The branch and bound algorithm used to find the optimal branching goes as follows. First an initial bound is established by connecting the minimum LBC split node containing all prizes to root. Then, the true cost is calculated as described in the problem definition, using \cref{eq:Tred}. After initializing the cost, potential trees are generated, starting from the root, adding valid edges to the tree as long as the total LBC does not exceed the current minimum TC. Three factors define valid edges that can be added to the tree. The first ensures that each prize is only collected once, using the running list of collected prizes to determine edges that connect nodes with no overlapping prizes. The second condition is that only edges whose parent nodes are currently in the tree may be added, ensuring the branching nature of the final result. The third condition is that, based on an arbitrary ordering of the edges, only higher numbered edges can be added. This condition avoids repetition in exploring trees that would be caused considering the order edges are added to the same parent node as unique trees. Once all prizes have been collected, the tree is considered complete, and the TC of the configuration is calculated using the proposed modeling technique. This TC is compared to the current minimum TC, and if lower, the minimum TC and optimal branching are updated. A detailed algorithm for the branch and bound technique is presented in \cref{alg:bnb}.

\begin{algorithm}
    \caption{Branch and bound for optimum branching}
    \label{alg:bnb}
    \begin{algorithmic}[1]
    \State{\textbf{inputs} $\mathcal{G}_{full}=(\mathcal{V},\mathcal{E}),Pr_s,LBC_{\mathcal{E}}, LBC_{s}$}
    \State{\textbf{global} $TC_{best},tr_{best}$}
    \State{$\varepsilon_{init} \gets \mathcal{E}(i)$ s.t. $\mathcal{E}^-_i= 0$ and $\text{card}\{Pr_s(\mathcal{E}^+_i)\} = n_u$}
    \State{$C_{init} \gets LBC_{\mathcal{E}}(\varepsilon_{init})+LBC_{s}(\mathcal{E}^+_i)$}
    \State{$\mathcal{G}_{init}\gets(\{v_0,\mathcal{E}^+_i\},\{\varepsilon_{init}(i)\})$ s.t. $C_{init}(i)=\min C_{init}$}
    \State{Find $TC$ for tree $\mathcal{G}_{init}$}
    \State{$TC_{best}\gets TC$}
    \State{$\mathcal{G}_{best}\gets \mathcal{G}_{init}$}
    \ForAll{$i$ s.t. $\mathcal{E}^-_i= 0$}
    \State{$\mathcal{G}_i\gets(\{v_0,\mathcal{E}^+_i\},\{\varepsilon_i\})$}
    \State{$Pr_i \gets Pr_s(\mathcal{E}^+_i)$}
    \State{$LBC_{run} \gets LBC_\mathcal{E}(\varepsilon_{i})+LBC_\mathcal{V}(\mathcal{E}^+_i)$}
    \State{\Call{GrowTree}{$\mathcal{G}_i,Pr_i,LBC_{run},i$}}
    \EndFor
    \Procedure{GrowTree}{$\mathcal{G},Pr,LBC,ii$}
    \If{$\text{card}\{Pr\}=n_u$ and $LBC<TC_{best}$}
        \State{Find $TC$ for tree $\mathcal{G}$}
        \If{$TC<TC_{best}$}
        \State{$TC_{best}\gets TC$}
        \State{$\mathcal{G}_{best}\gets \mathcal{G}$}
        \EndIf
    \ElsIf{$LBC<TC_{best}$}
        \ForAll{$j<ii$ \textbf{s.t.} $\mathcal{E}_j^-=\mathcal{E}_{\min ii}^-$ and $Pr_s(\mathcal{E}^+_j)\cap Pr=\varnothing$}
            \State{$\mathcal{G}_{new} \gets \mathcal{G}\cup \varepsilon_j$}
            \State{$Pr_{new} \gets Pr\cup Pr_s(\mathcal{E}^+_j)$}
            \State{$LBC_{new} \gets LBC+LBC_\mathcal{E}(e_{j})+LBC_\mathcal{V}(\mathcal{E}^+_i)$}
            \State{\Call{GrowTree}{$\mathcal{G}_{new}, Pr_{new}, LBC_{new}, \begin{bmatrix} ii & j \end{bmatrix}$}}
        \EndFor
        \ForAll{$j$ \textbf{s.t.} $\mathcal{E}_j^-=\mathcal{E}_{ii}^+$ and $Pr_s(\mathcal{E}^+_j)\cap Pr=\varnothing$}
            \State{$\mathcal{G}_{new} \gets \mathcal{G}\cup e_j$}
            \State{$Pr_{new} \gets Pr\cup Pr_s(\mathcal{E}^+_j)$}
            \State{$LBC_{new} \gets LBC+LBC_\mathcal{E}(e_{j})+LBC_\mathcal{V}(\mathcal{E}^+_i)$}
            \State{\Call{GrowTree}{$\mathcal{G}_{new}, Pr_{new}, LBC_{new}, j$}}
        \EndFor
    \EndIf
    \EndProcedure
    \end{algorithmic}
\end{algorithm}

\subsection{Results}
\label{sec:results}

\begin{table}
\caption{Results of the energy loss minimization problem.}
\label{tble:results}
\centering
\begin{tabular}{l c c}
\toprule
Layout& Length [m] & Enthalpy [W]\\
\midrule
    Length-Minimized & $1.26\times10^3$ & $1.63\times10^5$ \\
    Loss-Minimized & $1.60\times10^3$ & $1.39\times10^5$\\
\bottomrule
\end{tabular}
\end{table}

\begin{figure}
    \centering
    \includegraphics[width=3.33in]{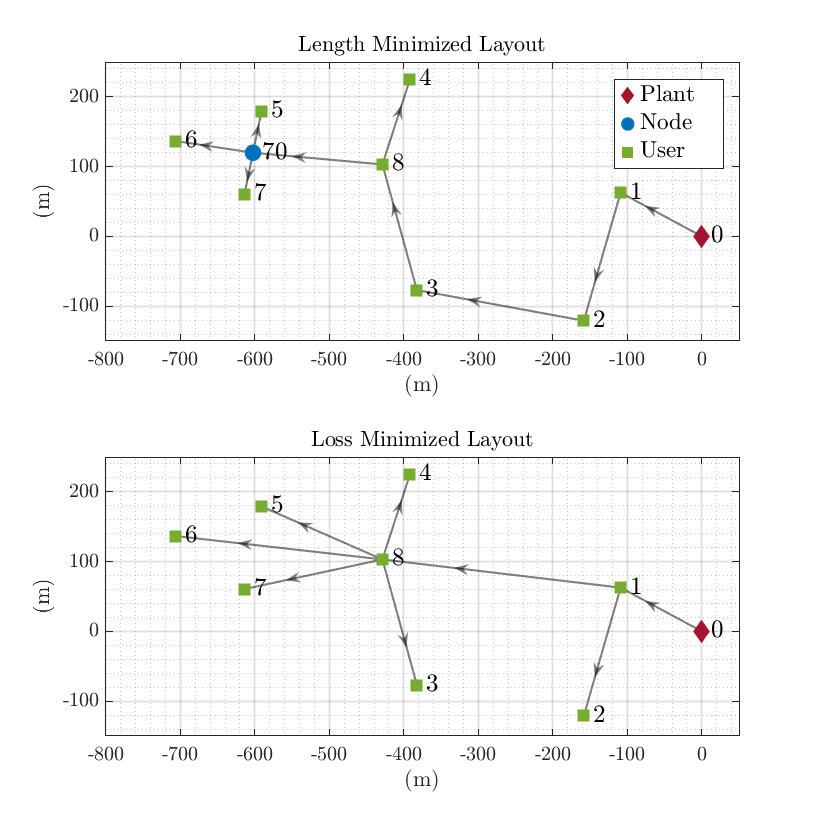}
    \caption{Resulting length optimized layout (left) and energy loss optimized layout (right).}
    \label{fig:results}
\end{figure}
The layouts resulting from minimizing the length of the pipes is compared to the layout resulting from minimizing the enthalpy losses in \cref{fig:results}. Additionally, the total length of each configuration and total enthalpy loss are provided for each layout in \cref{tble:results}. Overall, considering heat transfer during the design of the DHN layout resulted in a 15\% reduction in enthalpy losses as compared to the length minimized layout. The length minimized layout favors connecting multiple users to the same branch, while the enthalpy minimized layer results in more branches being created for individual users. This is due to the heat loss reduction of the reduced flow rate and smaller diameter pipes in branches with single users. Additionally, the length minimized layout uses a midpoint, to keep flow together and only split close to the user, while the enthalpy minimized layout favors splitting earlier, allowing for more smaller-diameter pipes to minimize the heat losses as the water is being delivered to the users. The candidate midpoint set contained over 1 million potential midpoints and the solution graph contained over 2.5 million edges. Using the developed modeling technique, all viable potential configurations were evaluated in around 44 hours.

\section{Conclusion}
\label{sec:conclusion}
This paper presents a novel set of algorithms needed to generate the state-space model of a branching multi-user DHN. The proposed model equations are validated using data collected from a lab-scale DHN considering a two-user layout. The calibrated model accurately captures both the fast and slow temperature dynamics observed in the network, with an nRMSE of 0.15 in return temperature. Additionally, the ability to automatically generate models of new network configurations is demonstrated through its application to the optimal design problem considering energy losses during steady-state peak network operations. The optimal design problem is formulated as an dynamic optimum branching problem and solved via a branch and bound algorithm. Over 2.5 million configurations are considered, and the resulting design reduces heat losses by 15\% as compared to the length minimized layout. \par
Future work will use this modeling framework to develop advanced model predictive controllers for large-scale DHNs. The flexibility of the modeling technique to new topology configurations will allow it to be used to rapidly generate models of a partitioned large-scale DHN for distributed control, and allow the plant model to easily adapt to any topology changes due to added users or network faults. Furthermore, the  dynamic optimum branching problem formulation and solution technique developed here can be adapted to consider the optimal placement of distributed renewable energy sources, a key decision when integrating low temperature heat sources in a DHN.   
\section*{Funding Data}
\noindent
This material is based upon work supported by the National Science Foundation Graduate Research Fellowship Program under Grant No.  DGE-1343012.

\begin{nomenclature}
\EntryHeading{Variables}
\entry{$T$}{Temperature}
\entry{$t$}{Time}
\entry{$\dot{m}$}{Mass flow rate}
\entry{$\rho$}{Density}
\entry{$V$}{Volume}
\entry{$c_p$}{Constant pressure specific heat capacity}
\entry{$\dot{Q}$}{Rate of heat transfer}
\entry{$h$}{Conductive heat transfer coefficient}
\entry{$A_s$}{Surface area}
\entry{$c$}{Temperature equation coefficient}
\entry{$\Delta P$}{Pressure drop}
\entry{$A_c$}{Cross sectional area}
\entry{$k$}{Concentrated pressure loss coefficient}
\entry{$\lambda$}{Distributed pressure loss coefficient}
\entry{$\zeta$}{Total pressure loss coefficient}
\entry{$L$}{Length}
\entry{$D$}{Diameter}
\entry{$a,b$}{Vector component of $A$ matrix}
\entry{$e$}{Vector component of $E$ matrix}
\entry{$n$}{number of elements}
\entry{$\mathcal{G}$}{Graph}
\entry{$\mathcal{V}$}{Set of nodes}
\entry{$\mathcal{E}$}{Set of edges}
\entry{$v$}{Single node of graph}
\entry{$\varepsilon$}{Single edge of graph}
\entry{$\alpha$}{Percent mass flow rate split}
\entry{$C$}{Cost}
\entry{$Pr$}{Prize set}
\entry{$LBC$}{Lower bound cost}
\entry{$TC$}{True cost}

\EntryHeading{Subscripts}
\entry{$in$}{Inlet}
\entry{$amb$}{Ambient}
\entry{$b$}{Building}
\entry{$0$}{Supply}
\entry{$S1, S2, S3$}{User supply segment}
\entry{$F$}{Feeding segment}
\entry{$R$}{Return segment}
\entry{$U$}{User without bypass}
\entry{$B$}{User with bypass}
\entry{$u$}{Users}
\entry{$s$}{Split nodes}
\entry{$p$}{Pipes}
\entry{$leaf$}{Leaf nodes}

\EntryHeading{Superscripts}
\entry{$-$}{Predecessor}
\entry{$+$}{Successor}
\entry{$\{i\}$}{Element number}
\end{nomenclature}

\bibliographystyle{asmejour}
\bibliography{sources}

\begin{thebibliography}{10}
\newcommand{\enquote}[1]{``#1''}
\providecommand{\url}[1]{\texttt{#1}}
\providecommand{\urlprefix}{}
\expandafter\ifx\csname urlstyle\endcsname\relax
  \providecommand{\doi}[1]{doi:\discretionary{}{}{}#1}\else
  \providecommand{\doi}{doi:\discretionary{}{}{}\begingroup
  \urlstyle{rm}\Url}\fi
\providecommand{\eprint}[2][]{\urlprefix\url{#1#2}}
\providecommand{\hrefurl}[2][]{\href{#1}{#2}}

\bibitem{delorenziSetupTestingSmart2020}
De~Lorenzi, A., Gambarotta, A., Morini, M., Rossi, M., and Saletti, C., 2020,
  \enquote{Setup and testing of smart controllers for small-scale district
  heating networks: {An} integrated framework,}
  \hrefurl{https://doi.org/10.1016/j.energy.2020.118054}{Energy}, \textbf{205},
  p. 118054.

\bibitem{potocnikMachinelearningbasedMultistepHeat2021}
Potočnik, P., Škerl, P., and Govekar, E., 2021,
  \enquote{Machine-learning-based multi-step heat demand forecasting in a
  district heating system,}
  \hrefurl{https://www.sciencedirect.com/science/article/pii/S0378778820334599}{Energy
  and Buildings}, \textbf{233}, p. 110673.

\bibitem{xueMachineLearningbasedLeakage2020}
Xue, P., Jiang, Y., Zhou, Z., Chen, X., Fang, X., and Liu, J., 2020,
  \enquote{Machine learning-based leakage fault detection for district heating
  networks,} \hrefurl{https://doi.org/10.1016/j.enbuild.2020.110161}{Energy and
  Buildings}, \textbf{223}, p. 110161.

\bibitem{wangNewModelOnsite2018}
Wang, H., Meng, H., and Zhu, T., 2018, \enquote{New model for onsite heat loss
  state estimation of general district heating network with hourly
  measurements,}
  \hrefurl{https://doi.org/10.1016/j.enconman.2017.11.062}{Energy Conversion
  and Management}, \textbf{157}, pp. 71--85.

\bibitem{zhaoInfluencingParametersAnalysis2019}
Zhao, J. and Shan, Y., 2019, \enquote{An Influencing Parameters Analysis of
  District Heating Network Time Delays Based on the {CFD} Method,}
  \hrefurl{https://doi.org/10.3390/en12071297}{Energies}, \textbf{12}(7), p.
  1297.

\bibitem{betancourtschwarzModifiedFiniteVolumes2019}
Betancourt~Schwarz, M., Mabrouk, M.~T., Santo~Silva, C., Haurant, P., and
  Lacarrière, B., 2019, \enquote{Modified finite volumes method for the
  simulation of dynamic district heating networks,}
  \hrefurl{https://doi.org/10.1016/j.energy.2019.06.038}{Energy}, \textbf{182},
  pp. 954--964.

\bibitem{vanderheijdeDynamicEquationbasedThermohydraulic2017}
van~der Heijde, B., Fuchs, M., Ribas~Tugores, C., Schweiger, G., Sartor, K.,
  Basciotti, D., Müller, D., Nytsch-Geusen, C., Wetter, M., and Helsen, L.,
  2017, \enquote{Dynamic equation-based thermo-hydraulic pipe model for
  district heating and cooling systems,}
  \hrefurl{https://doi.org/10.1016/j.enconman.2017.08.072}{Energy Conversion
  and Management}, \textbf{151}, pp. 158--169.

\bibitem{salettiControlorientedScalableModel2022}
Saletti, C., Zimmerman, N., Morini, M., Kyprianidis, K., and Gambarotta, A.,
  2022, \enquote{A control-oriented scalable model for demand side management
  in district heating aggregated communities,}
  \hrefurl{https://doi.org/10.1016/j.applthermaleng.2021.117681}{Applied
  Thermal Engineering}, \textbf{201}, p. 117681.

\bibitem{vandermeulenSimulationbasedEvaluationSubstation2020}
Vandermeulen, A., Van~Oevelen, T., van~der Heijde, B., and Helsen, L., 2020,
  \enquote{A simulation-based evaluation of substation models for network
  flexibility characterisation in district heating networks,}
  \hrefurl{https://doi.org/10.1016/j.energy.2020.117650}{Energy}, \textbf{201},
  p. 117650.

\bibitem{feltenIntegratedModelCoupled2020}
Felten, B., 2020, \enquote{An integrated model of coupled heat and power
  sectors for large-scale energy system analyses,}
  \hrefurl{https://doi.org/10.1016/j.apenergy.2020.114521}{Applied Energy},
  \textbf{266}, p. 114521.

\bibitem{huoOperationOptimizationDistrict2022}
Huo, S., Wang, J., Qin, Y., and Cui, Z., 2022, \enquote{Operation optimization
  of district heating network under typical modes for improving the economic
  and flexibility performances of integrated energy system,}
  \hrefurl{https://doi.org/10.1016/j.enconman.2022.115904}{Energy Conversion
  and Management}, \textbf{267}, p. 115904.

\bibitem{simonssonReducedOrderModelingThermal2022}
Simonsson, J., Atta, K., and Birk, W., 2022, \enquote{Reduced-Order Modeling of
  Thermal Dynamics in District Energy Networks Using Spectral Clustering,}
  \textit{2022 IEEE Conference on Control Technology and Applications (CCTA)},
  Trieste, Italy, August 23--25, pp. 144--150,
  \doi{10.1109/CCTA49430.2022.9965996}.

\bibitem{guelpaCompactPhysicalModel2019}
Guelpa, E. and Verda, V., 2019, \enquote{Compact physical model for simulation
  of thermal networks,}
  \hrefurl{https://doi.org/10.1016/j.energy.2019.03.064}{Energy}, \textbf{175},
  pp. 998--1008.

\bibitem{wackEconomicTopologyOptimization2023}
Wack, Y., Baelmans, M., Salenbien, R., and Blommaert, M., 2023,
  \enquote{Economic topology optimization of District Heating Networks using a
  pipe penalization approach,}
  \hrefurl{https://doi.org/10.1016/j.energy.2022.126161}{Energy}, \textbf{264},
  p. 126161.

\bibitem{suOptimizingPipeNetwork2022}
Su, L., Nie, T., On~Ho, C., Yang, Z., Calvez, P., Jain, R.~K., and Schwegler,
  B., 2022, \enquote{Optimizing pipe network design and central plant
  positioning of district heating and cooling System: A Graph-Based
  Multi-Objective genetic algorithm approach,}
  \hrefurl{https://doi.org/10.1016/j.apenergy.2022.119844}{Applied Energy},
  \textbf{325}, p. 119844.

\bibitem{williamsDynamicalGraphModels2017}
Williams, M.~A., Koeln, J.~P., Pangborn, H.~C., and Alleyne, A.~G., 2017,
  \enquote{Dynamical Graph Models of Aircraft Electrical, Thermal, and
  Turbomachinery Components,}
  \hrefurl{https://doi.org/10.1115/1.4038341}{Journal of Dynamic Systems,
  Measurement, and Control}, \textbf{140}(4), p. 041013.

\bibitem{jogwarDistributedControlArchitecture2019}
Jogwar, S.~S., 2019, \enquote{Distributed control architecture synthesis for
  integrated process networks through maximization of strength of
  input–output impact,}
  \hrefurl{https://doi.org/10.1016/j.jprocont.2019.08.009}{Journal of Process
  Control}, \textbf{83}, pp. 77--87.

\bibitem{salettiDevelopmentAnalysisApplication2020}
Saletti, C., Gambarotta, A., and Morini, M., 2020, \enquote{Development,
  analysis and application of a predictive controller to a small-scale district
  heating system,}
  \hrefurl{https://doi.org/10.1016/j.applthermaleng.2019.114558}{Applied
  Thermal Engineering}, \textbf{165}, p. 114558.

\bibitem{nussbaumerHandbookPlanningDistrict2020}
Nussbaumer, T., Thalmann, S., Jenni, A., and Ködel, J., 2020,
  \textit{\hrefurl{https://www.verenum.ch/Dokumente/Handbook-DH_V1.0.pdf}{Handbook
  on Planning of District Heating Networks}}, Swiss Federal Office of Energy.

\bibitem{bohmSimpleModelsOperational2002}
B{\o}hm, B., Ha, S.-k., Kim, W.-t., Kim, B.-k., Koljonen, T., Larsen, H.~V.,
  Lucht, M., Park, Y.-s., Sipil{\"a}, K., Wigbels, M., et~al., 2002,
  \enquote{Simple models for operational optimisation,} Contract,
  \textbf{524110}(2002,S1).

\bibitem{blizardDynamicallySimilarLabscale2023}
Blizard, A. and Stockar, S., 2023, \enquote{A dynamically similar lab-scale
  district heating network via dimensional analysis,}
  \hrefurl{https://doi.org/10.1016/j.enconman.2023.117446}{Energy Conversion
  and Management}, \textbf{293}, p. 117446.

\bibitem{piroutiEnergyConsumptionEconomic2013}
Pirouti, M., Bagdanavicius, A., Ekanayake, J., Wu, J., and Jenkins, N., 2013,
  \enquote{Energy consumption and economic analyses of a district heating
  network,} \hrefurl{https://doi.org/10.1016/j.energy.2013.01.065}{Energy},
  \textbf{57}, pp. 149--159.

\bibitem{cimbala2006fluid}
Cimbala, J.~M. and Cengel, Y.~A., 2006, \textit{Fluid mechanics: fundamentals
  and applications}, McGraw-Hill Higher Education.

\bibitem{gabrielaitieneModellingTemperatureDynamics2007}
Gabrielaitiene, I., Bøhm, B., and Sunden, B., 2007, \enquote{Modelling
  temperature dynamics of a district heating system in Naestved, Denmark—A
  case study,} \hrefurl{https://doi.org/10.1016/j.enconman.2006.05.011}{Energy
  Conversion and Management}, \textbf{48}(1), pp. 78--86.

\bibitem{anconaApplicationDifferentModeling2019}
Ancona, M.~A., Branchini, L., De~Lorenzi, A., De~Pascale, A., Gambarotta, A.,
  Melino, F., and Morini, M., 2019, \enquote{Application of different modeling
  approaches to a district heating network,}
  \hrefurl{https://doi.org/10.1063/1.5138742}{{AIP} Conference Proceedings},
  \textbf{2191}(1), p. 020009, Publisher: American Institute of Physics.

\bibitem{masatinEvaluationFactorDistrict2016}
Masatin, V., Latõšev, E., and Volkova, A., 2016, \enquote{Evaluation Factor
  for District Heating Network Heat Loss with Respect to Network Geometry,}
  \hrefurl{https://doi.org/10.1016/j.egypro.2016.09.069}{Energy Procedia},
  \textbf{95}, pp. 279--285.

\bibitem{edmonds1967optimum}
Edmonds, J. et~al., 1967, \enquote{Optimum branchings,} Journal of Research of
  the national Bureau of Standards B, \textbf{71}(4), pp. 233--240.

\bibitem{manerbaTravelingPurchaserProblem2017}
Manerba, D., Mansini, R., and Riera-Ledesma, J., 2017, \enquote{The Traveling
  Purchaser Problem and its variants,}
  \hrefurl{https://doi.org/10.1016/j.ejor.2016.12.017}{European Journal of
  Operational Research}, \textbf{259}(1), pp. 1--18.

\end{thebibliography}
\end{document}